\documentclass[%
 reprint,
 amsmath,amssymb,
 aps,
]{revtex4-2}

\usepackage{graphicx}% Include figure files
\usepackage{dcolumn}% Align table columns on decimal point
\usepackage{bm}% bold math
\usepackage{hyperref}

\usepackage{aas_macros}
\usepackage[capitalize]{cleveref}
\usepackage[dvipsnames,table]{xcolor}
\hypersetup{urlcolor=RedViolet,
	    citecolor=ForestGreen ,
	    linkcolor=RoyalBlue, 
	    colorlinks=true}
\usepackage[normalem]{ulem}

\usepackage{soul}

\newcommand{\Mpch}{$h^{-1}$Mpc}
\newcommand{\kpch}{$h^{-1}$kpc}
\newcommand{\msunh}{$h^{-1}M_{\odot}$}

\newcommand{\ie}{\textit{i.e.}}

\newcommand{\elephant}{\textsc{elephant}}

\newcommand{\lcdm}{$\mathrm{\Lambda \text{CDM}}$}

\newcommand{\pk}{$P(k)$}

\newcommand{\tweb}{\textsc{t-web}}

\begin{document}

\preprint{APS/123-QED}

\title{Bending the web: exploring the impact of modified gravity on the density field and halo properties within the cosmic web}% Force line breaks with \\

\author{Suhani Gupta$^{1}$}
\thanks{email: gupta@cft.edu.pl}

\author{Simon Pfeifer$^{2}$}
\author{Punyakoti Ganeshaiah Veena$^{3,4}$}
\author{Wojciech A. Hellwing$^{1}$}

\affiliation{$^{1}$Center for Theoretical Physics, Polish Academy of Sciences, Al. Lotnik\'ow 32/46, 02-668 Warsaw, Poland}%

\affiliation{$^{2}$Leibniz-Institut f\"ur Astrophysik Potsdam, An der Sternwarte 16, D-14482 Potsdam, Germany}

\affiliation{$^{3}$Department of Physics, Technion, Haifa 3200003, Israel}
\affiliation{$^{4}$Department of Physics, University of Genova, Via Dodecaneso, 33, Genoa, Italy}

\date{\today}

\begin{abstract}

This work investigates the impact of different Modified Gravity (MG) models on the large-scale structures (LSS) properties in relation to the cosmic web (CW), using N-body simulations of $f(R)$ and the normal branch of Dvali-Gabadadze-Porrati (nDGP) models. We analyse the impact of the MG effect on the density field through density distribution and clustering statistics, and assess its influence on halo properties by examining the halo mass function and spin. We find that the Probability Density Function (PDF) of dark matter density fields shift towards lower densities for stronger variants of $f(R)$ and nDGP. Additionally, when segregated into CW environments, the stronger variants show a higher mean density in knots, and a lower mean density in voids compared to \lcdm. For higher-order clustering statistics relative to \lcdm, the scale-dependent $f(R)$ variants exhibit a greater non-monotonic deviation as a function of scale when segregated into environments, compared to nDGP. Additionally, the halo mass function separated into CW environments shows a similar behaviour, introducing complex trends as a function of mass for $f(R)$ and nDGP models. We also report up to a $\sim$15\% enhancement in the angular momentum of halos in $f(R)$ gravity models compared to \lcdm{}, with similar differences when considering environmental segregation. We demonstrate that this difference in the spin arises largely due to different tidal torquing across the various MG models. Therefore, studying higher-order statistics of the cosmological fields and halo properties separated into CW components probes the additional physics contained within the MG models. We conclude that considering the effect of CW in MG studies increases the constraining power of these LSS statistics, and can further aid the distinction between the cosmologies that have an identical expansion history to the standard \lcdm{} but differing underlying physics, such as the MG models presented in this work.

\end{abstract}

\maketitle

\section{Introduction}

The standard cosmological paradigm, based on Einstein’s theory of General Relativity (GR) and known as \lcdm{}, has proven to be a highly successful model for describing the evolution of the Universe. This standard cosmological model explains very well the light element abundance from the primordial nucleosynthesis, the temperature and polarisation anisotropies of the cosmic microwave background radiation, the large-scale clustering of matter using multiple probes, as well as the late-time accelerated expansion of the Universe \citep{bbn_review,Planck2018,sdss4_highz,sdss2_snls_de,des_y1_prl}. However, despite its observational success, the standard model carries its own set of challenges. Within the \lcdm{} model, the composition of the matter content in the Universe is dominated by \textit{dark matter} (DM). Moreover, the observed cosmic acceleration is governed by a mysterious \textit{dark energy} component, and the $\Lambda$ (cosmological constant) here is the leading candidate to explain this cosmic speed-up. The physical nature of both these components, despite their dominant composition, remains unknown. Furthermore, to explain the formation of small initial perturbations that set the seeds for structure formation, the theory of \textit{inflation} is invoked which is still far from being observationally tested. 

The phenomenological nature of \lcdm{}, along with the theoretical and observational challenges it confronts \cite{cc_problem,lcdm_problems1,lcdm_small_scale_challenges,lcdm_tensions_sn_quasar_grb}, has driven the search for alternative scenarios or extensions to the concordance model. One active avenue focuses on attributing the late-time accelerated expansion to beyond-GR extensions (typically scalar-tensor theories) rather than to the vanishingly small $\Lambda$. Such models are termed as ``Modified Gravity''(MG). These MG models are constructed in such a way that they have negligible consequences at early times and share the same expansion history and cosmological background as \lcdm{}. The effect of these MG models is incorporated in the perturbation equations that govern the gravitational dynamics of large-scale structures (LSS). Hence, we would expect that the observables and measures associated with LSS formation and evolution are a promising probe to further investigate departures from the standard GR predictions.

In this work, we focus on families of two such MG phenomenologies: namely $f(R)$ \citep{fR_gravity_theories,HS_fR_2007}, and the normal branch of Dvali-Gabadadze-Porrati (nDGP) gravity models \cite{ndgp_2000,ndgp_2002}. These MG theories offer a perfect test bed to explore the freedom of modifying the Einstein-Hilbert action, in order to produce a physical mechanism effectively mimicking the action of $\Lambda$, that would result in cosmic acceleration. In the $f(R)$ model, we consider the possibility of extra non-linear function $f(R)$ of the Ricci scalar $R$ in the Einstein-Hilbert action. This results through additional coupling between matter and scalar field. In this model, the accelerated expansion of the Universe is produced by this $f(R)$ term, thereby replacing $\Lambda$ in the action integral. The second MG model we consider is nDGP model introduced in \cite{ndgp_2000,ndgp_2002}. nDGP is one of the simplest and most popular explanations of the accelerated expansion of the Universe through higher dimensional spacetime. In this gravity model, all standard forces, except gravity, are confined on a four-dimensional brane that is embedded in five-dimensional bulk spacetime, and gravity propagates in an additional fifth-dimension. Here, the scalar is identified as the brane-bending mode which describes the deformation of the 4D brane in the 5D bulk spacetime, and the displacement of the scalar induces the additional \textit{fifth-force} for this MG variant.

Most of the viable MG models invoke \textit{screening mechanisms} to suppress the effect of the \textit{fifth-force} in regimes with strong constraints on gravity. Both $f(R)$ and nDGP can be divided into two general categories, depending on the physical mechanism of the \textit{fifth-force} screening they invoke. We work with the functional form of $f(R)$ gravity described in \cite{HS_fR_2007}, which implements the Chameleon screening mechanism \citep{Khoury2003PRD}, and the nDGP model implements the Vainshtein screening mechanism \cite{Vainshtein_1972}. These two ways of suppressing the \textit{fifth-force} are both extremely non-linear and have fundamental physical differences. The Chameleon mechanism makes the scalar field significantly massive in high-density regions by inducing an effective Yukawa-like screening. The effectiveness of the Chameleon depends on the local density thereby inducing environmental effects in the resultant \textit{fifth-force} dynamics. The Vainshtein screening mechanism, on the other hand, makes the scalar field kinetic terms very large in the vicinity of massive bodies and as a result, the scalar field decouples from matter resulting in the screening of the \textit{fifth-force}. Vainshtein screening depends only on the mass and distance from a body, and shows no explicit dependence on the cosmic environment. An elaborate description of these screening mechanisms is discussed in e.g.   \citep{screening_review,Khoury2003PRD,HS_fR_2007,Vainshtein_1972,nonlinear_interactions_nDGP,Brax_screening_mg,astrophysical_tests_screening,cosmological_tests_mg}.

Analytic predictions and cosmological simulations have shown that beyond the linear growth of density perturbations, complex patterns and structures emerge in the DM density fields at large scales. The LSS beyond tens of Mpc is manifested in the form of web-like structures, called the \textit{cosmic web} (CW) \cite{cw_bond_1996}. 
The existence of the CW is a consequence of the anisotropic collapse of the cosmic overdensity, which is due to gravitational instability \citep{Zeldovich_1970,peebles_1980,bond_1996,bond_myers_1996,clusters_CW,Millennium1,HM_CW}. Once the gravitational clustering process begins to go beyond the linear growth phase, we see the emergence of complex patterns and structures in the density field. Under the effect of gravity, these structures evolve and cluster from tiny density and velocity perturbations in the early Universe, resulting in the formation of elongated structures called \textit{filaments}, or flattened structures called \textit{sheets/walls}. The filaments are connected by high-density \textit{knots/nodes}, and the walls surround large underdensities, or the \textit{voids}. Hence, it is expected that any modifications to the gravity theory, and consequently, to the underlying cosmological model, would significantly influence the evolution of the CW environment and the associated statistical properties \cite{Mnu_clustering_halos_voids,tidal_env_GAMA,GIGANTIS}.

%Impact of cw and MG on LSS properties
%CW influences the properties of halos and galaxies
The CW structure serves as an environment for the DM halos, and in-turn galaxies, to form and evolve. The dependence of DM halo properties on CW environment has been widely studied in the literature. Many works have shown systematic dependencies of the large-scale properties on the environment that hosts them \cite{hahn_cw_2007,evolution_cw_nexus,gama_filaments_voids,sdss_cw,abundance_cw_2,halo_abundance_cw,cosmic_ballet_1,cosmic_ballet_2,cosmic_ballet_3,cosmic_ballet_1,cosmic_ballet_2,cosmic_ballet_3,baryons_cw_illustrisTNG,cw_eagle,cw_pk_1,cw_pk_2,hellwing_CW,conc_spin_shape_CW}. While substantial efforts are being made to link the properties of the CW elements to the LSS they host, there is currently a limited understanding about the information these environments carry of the underlying cosmological model. Moreover, there have been limited studies that consider the role of MG in the context of CW environments. The influence of $f(R)$ gravity on halo characteristics in various CW environments has been examined in a study by \cite{hmf_fR_cw}, while the effect of $f(R)$ gravity on filamentary structures within the CW is explored in \cite{massive_gravity_CW}. Moreover, in \cite{CW_FDM}, the authors show how distinct DM models impact both DM density and halo attributes across diverse CW environments.

%Motivation: lack of combined study and more information
Examining the influence of MG on the large-scale properties within varied CW environments is significant for several reasons: \textit{Firstly,} the CW can be broadly classified on the basis of the density and velocity fields \cite{tweb,vweb_yehuda,nexus} that are sensitive to modifications to the underlying gravitational forces (\cite{pkmg_sg,hellwing_velocities} and the references within). Consequently, we can expect that the statistics pertaining the LSS would provide more information when studied separately for each CW environment. These studies could break degeneracies, and improve the constraints on the cosmological parameters through the analysis of the cosmological information regarding the matter distribution in different cosmological environments \cite{cw_pk_1,cw_pk_2,rsd_split_density}. \textit{Secondly,} CW marks a transition of the density field from a primordial Gaussian random field, to highly non-linear density field that collapses to form cosmic structures, like halos. 
The weakly nonlinear CW comprises features on the scales of tens of Mpc, in which large structures have not lost memory of the nearly homogeneous primordial state from which they formed, and provide a direct link to early Universe physics. This shows the significant influence of these intermediate scales where additionally we do not have enough constraints on the theory of gravity. Moreover, it is on these intermediate scales where the CW largely manifests. As a result, CW studies in MG theories further provide a potential tool to test alternate structure formation scenarios to GR, and in-turn to constrain gravity. \textit{Thirdly}, screening is the least effective in low-densities, making voids a perfect test-beds to test for MG signatures \cite{fR_void_1,ndgp_void_2,chameleon_voids}. \textit{Lastly}, studying the DM and halo properties in different CW environments can help draw a boundary between the halo properties that are affected, and not affected by screening. Considering each environment separately could help us to individually examine the LSS results in both the screened and unscreened regions, rather than averaging the effects over all the environments. This will further help us distinguish the screening physics from the impact of the \textit{fifth-force}.

%Structure of the article
Motivated by the factors mentioned above, this work explores the impact of MG models on various LSS properties in different CW environments. The differences in densities, tidal forces, and anisotropies inherent to each CW environment, coupled with the influence of the \textit{fifth-force} and the screening mechanisms, predominantly contribute to these distinctions. Here, we first describe our simulations and datasets in \cref{sec:sims}, and our CW classification in \cref{sec:cw_classification}. We quantify the MG impact on DM density fields and the associated statistics across CW environments in \cref{subsec:pdf,subsec:cumulants}, and then we study how differently halo properties in each CW environment are impacted by the MG dynamics in \cref{subsec:hmf,subsec:spin}. We conclude this work with the summary and discussion in \cref{sec:summary_discussions}.

\section{Simulations and data-sets}
\label{sec:sims}
In this work, we use the \elephant{} (Extended LEnsing PHysics using ANalytic ray Tracing) suite of MG N-body simulations \cite{ALAM2020_ELEPHANT}. For these simulations, we have the Hu-Sawicki (HS) form of $f(R)$ gravity \cite{HS_fR_2007}, which implements Chameleon screening \cite{Khoury2003PRD}, and the Vainshtein screening mechanism \cite{Vainshtein_1972} is invoked by the normal branch of the Dvali-Gabadadzi-Porrati (nDGP) model \citep{ndgp_2000,ndgp_2002} to  satisfy the high-density solar system constraints. For more details of these models and simulations, we refer the reader to  \cite{ALAM2020_ELEPHANT,hmf_sg,pkmg_sg,hierarchical_fR,MG_SIGNATURES_HIERARCHICAL_CLUSTERING,rsd_bias,drozda_2022}. As both $f(R)$ and nDGP models deviate negligibly from \lcdm{} at early times, the simulations for all the cosmologies are started using the same initial conditions which also helps us to avoid any anomalies that could arise due to differences in phases of the initial density fields. As a result, any difference in the clustering dynamics can be attributed directly to the modified gravitational evolution.

HS $f(R)$ simulations are run using the \textsc{ecosmog} code \cite{ECOSMOG_1,ECOSMOG_2}, and the nDGP simulations are run using the \textsc{ecomog-v} code \cite{ECOSMOG_V_1, speeding_vainshtein}. Both are parallelised Adaptive-mesh-refinement (AMR) codes for MG, and are based on the publicly available \textsc{ramses} code \cite{RAMSES_2002}. The \elephant{} simulations consider $\Lambda$CDM as the fiducial model, 2 variants of HS $f(R)$ gravity with model parameters: $n=1$ and $|f_{R0}| = 10^{-5}$ and $10^{-6}$ (decreasing order of variation from \lcdm{}) which are here referred to as F5 and F6 respectively. For the nDGP model, we have two variants, with the model parameter r$_{c}$H$_{0}$ = 1, and 5 (again with decreasing order of variation from \lcdm{}) written as N1 and N5 respectively.
%\PGVc{I would put the sentence for more details of the model.. here.}

The \textsc{elephant} simulations were run with 1024$^{3}$ collision-less DM particles, each of mass $m_{p} = 7.798 \times 10^{10}$ \msunh, with a  comoving force resolution $\epsilon = 15$ \kpch{} in a box of size 1024 \Mpch{}. The simulation are run from an initial redshift $z_{ini}$ = 49, to a final redshift $z_{final}$ = 0, with the initial conditions generated by \textsc{MPgrafic} \cite{MPGraphic}, using the Zel'dovich Approximation \cite{Zeldovich_1970}. A high value for the initial redshift is required as a consequence of employing first-order Lagrangian perturbation theory \cite{LPT_TRANSIENT_SCOCCIMARO}. Furthermore, to reduce the effect of sample variance, each model has $5$ independent realizations. The background cosmological parameters of the simulations are: $\Omega_{m}= 0.281$, $\Omega_{b} = 0.046$, $\Omega_{\text{CDM}} = 0.235$, $\Omega_{\Lambda} = 0.719$, $\Omega_{\nu} = 0$, $H/100 = h = 0.697$ km $s^{-1} Mpc^{-1}$, $n_{s} = 0.971$ and $\sigma_{8} = 0.842$. These background parameters apply to the cosmological background in both \lcdm{} and MG cosmologies.

Halos are identified using the Robust Overdensity Calculation using a K-Space Topologically Adaptive Refinement \textsc{(rockstar)} halo finder \cite{ROCKSTAR}, which uses
%\spcc{Do you mean "evolves" or identifies, or something else?} 
the DM particles in six-dimensional phase-space to identify halos. The halo finder first identifies 3D overdensities using a Friend-of-Friend (FOF) method to assign particles to the halos. It then generates a hierarchy of these FoF subgroups by assigning the particles based on an adaptive linking length in 6D phase space. The entire process is parallelised, making it highly efficient. The mass of a halo, $M_{200}$, is defined as the mass contained in a spherical region of radius $R_{200}$, with a mean density equal to  $200$ times the critical density of the Universe, $\rho_{crit}(z) \approx 2.775 \times 10^{11} h^{2} M_{\odot}Mpc^{-3}$.%\PGVc{Is it critical or mean density? Maybe you checked, but just to be sure.}
We restrict our analysis to halos with mass equal to or greater than 100 times $m_{p}$ to minimise the shot noise or resolution effects. In principle, the presence of the \textit{fifth-force} would require a modification to \textsc{rockstar} (or in general to any standard halo-finding algorithm). Nonetheless, the authors in \cite{sf_f5} found the effect of this modification to be quite small, and hence we use the standard \textsc{rockstar} algorithm on both \lcdm{} and MG models.

\subsection{T-WEB classification of the cosmic web}
\label{sec:cw_classification}

A variety of different methods have been devised to classify the CW (e.g. \cite{tweb,spine_cweb,vweb_yehuda,nexus, cows}). These classifications are based on the morphological or the local geometric information of the density, tidal or velocity shear fields. The basis of quantifying the LSS using fields can be traced to the Zel'dovich Approximation (ZA) \cite{Zeldovich_1970}. The ZA played a key role in developing a view of the structure formation in which \textit{pancakes (or walls/sheets)} form after the collapse of matter along one axis, \textit{filaments} after collapse along a second axis, and \textit{nodes (or knots)} after collapse along a third axis. In the pioneering work of \cite{cw_bond_1996}, the authors further showed that the formation and the dynamical evolution of the CW is indeed a result of the tidal force fields. Building on these works, the authors in \cite{hahn_cw_2007} developed a classification scheme which is based on the signature of the tidal tensor, hence called the \tweb{}.

The \tweb{} is calculated via the Hessian of the gravitational potential $\phi$, the tidal shear tensor $T$, defined as 
\begin{equation}
    \label{eqn:hessian_tweb}
    T_{\alpha\beta}(\vec{x}, t) = \frac{\partial^2 \phi(\vec{x}, t)}{\partial x_{\alpha} \partial x_{\beta}}.
\end{equation}
The physical gravitational potential has been normalised by $4\pi G \bar{\rho}$ so that $\phi(\vec{x}, t)$ satisfies the Poisson equation $ \nabla^{2}\phi(\vec{x}, t) = \delta(\vec{x}, t)$. $\delta(\vec{x}, t)$ is the dimensionless matter overdensity, given by $\delta(\vec{x}, t) = \frac{\rho(\vec{x}, t)}{\overline{\rho}}-1 $, which quantifies local departure of the density field $\rho(\vec{x}, t)$ from the average uniform density field $\overline{\rho}$, at a given position $\vec{x}$, and for a specific time, $t$. In practice, the Hessian, which encodes the local variations in the density field, is calculated on a regular grid of the matter density field extracted from the simulation particle distribution via the Triangular-Shaped Cloud (TSC) assignment scheme. For this work, the DM density grid has $256^3$ voxels, with a size of 4 \Mpch{} per voxel. Furthermore, each voxel of the simulation box is categorised in different CW environments using the \tweb{} approach by smoothing the density fields at the resolution of the grid.

This tidal tensor can be represented by a real symmetric $3 \times 3$ matrix with eigenvalues $\lambda_1 > \lambda_2 > \lambda_3$. The Poisson equation is solved in Fourier space to obtain the potential over each grid cell, and the shear tensor is computed to obtain the corresponding eigenvalues. The number of eigenvalues larger than a threshold, $\lambda_{th}$, defines the CW environment: \textbf{\textcolor{blue}{knots}} (3 eigenvalues larger than $\lambda_{th}$), \textbf{\textcolor{YellowOrange}{filaments}} (2), \textbf{\textcolor{OliveGreen}{sheets}} (1) and \textbf{\textcolor{red}{voids}} (0). In the \tweb{} approach, $\lambda_{th}$ is an arbitrary quantity, which is usually considered order unity. In our work, we set $\lambda_{th} = 0.2$. This choice is motivated by previous studies that aimed to capture the visual impression of the CW \cite{tweb,tracing_cw_libeskind_2018}. We also use to the same threshold to define the CW elements in our MG models in order to maintain consistency for comparative analysis with the standard \lcdm{} baseline. Furthermore, $\lambda_{th}=0.2$ produces similar volume fractions for all the CW environments in each cosmology, thereby recovering the similar LSS in all models.

\section{Results}
\subsection{Probability Density Function}
\label{subsec:pdf}
\begin{figure}
    \centering
    \includegraphics[width=\columnwidth]{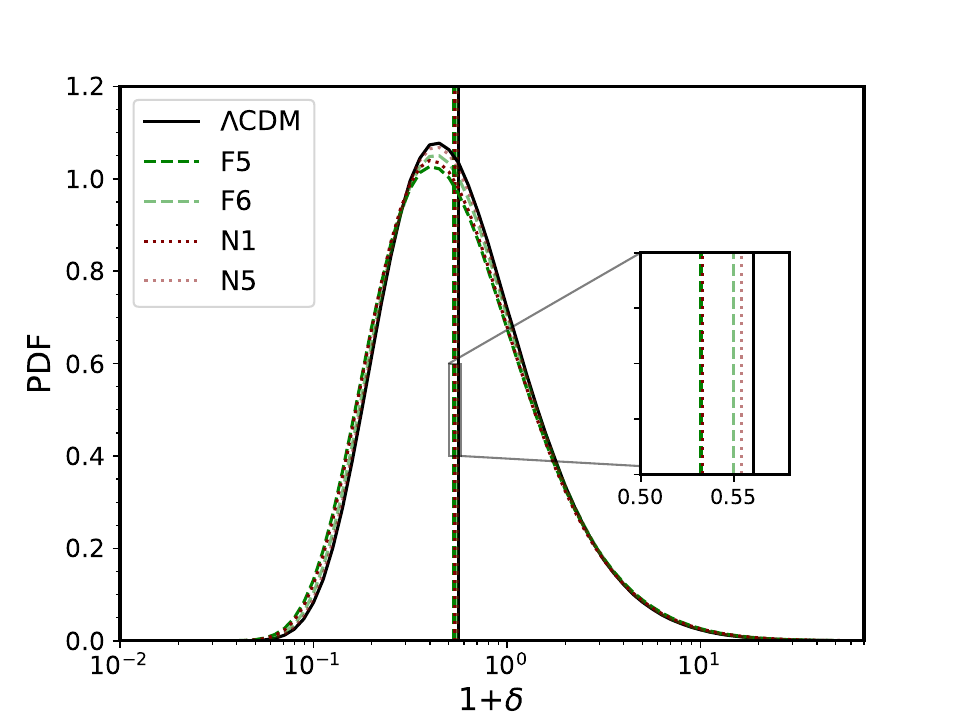}

    \caption{The PDF of matter overdensities in different cosmologies compared against \lcdm{}. This distribution is obtained using the density fields on a regular grid with a spacing of 4 \Mpch. The vertical lines are the median of the overdensity distribution for each model.}
    \label{fig:distribution_all}
\end{figure}

\begin{figure}
    \centering
    \includegraphics[width=\columnwidth]{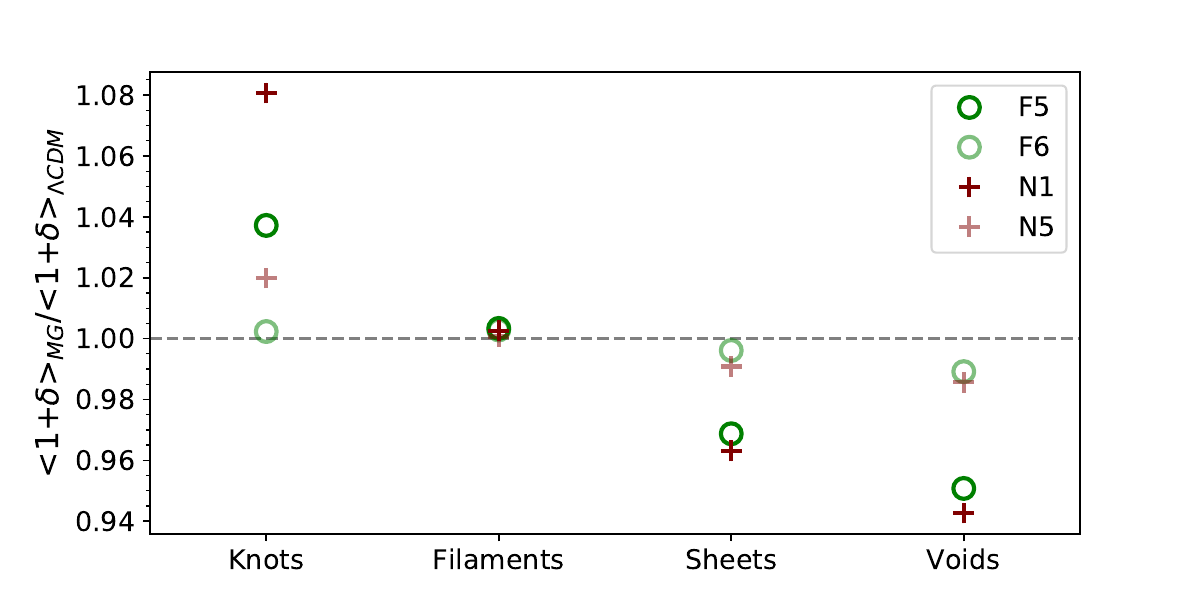}
                
    \caption{ The ratio of the mean overdensity in each MG model $w.r.t.$ \lcdm{}, separately for all CW environments.}
    \label{fig:mean_distribution_cw}
\end{figure}

The Probability Density Function (PDF) of matter over-densities can be approximated by a log-normal distribution in the standard \lcdm{} paradigm \cite{LN_coles_jones}. In this work, we study the distribution of the log of the over-densities, log(1+$\delta$), in different cosmologies. In \cref{fig:distribution_all} we plot this distribution for F5, F6, N1 and N5 against \lcdm. The PDF for all MG models is similar to \lcdm{} as it follows the log-normal distribution, with the mean of the overall distribution for all models centred at unity. However, when compared to \lcdm{}, we detect a shift in the peak as well as the median of the density distribution towards lower densities. This shift is larger for stronger variants F5 and N1, compared to their respective weaker counterparts F6 and N5 respectively.

To further visualise the effect of MG on the density distribution across different CW environments, we plot the mean of the density PDF for each CW environment relative to \lcdm{} in \cref{fig:mean_distribution_cw}. %The first column shows the mean, relative to \lcdm{}, for the entire density distribution which are also shown as vertical lines in \cref{fig:distribution_all}. 
For the case of the high density knots, MG variants have an enhancement in the mean density compared to \lcdm{}. The strong N1 model has the maximum enhancement in the mean, followed by the strong $f(R)$ variant F5, and then by the weaker N5 and F6 variants. As we move to the less dense environments, we see a systematic decrease in this ratio. The mean of the densities for all environments coincides with the \lcdm{} value for filaments. In sheets and voids, the stronger F5 and N1 gravity variants have a greater decrease in the density $w.r.t.$ \lcdm{} than the weaker F6 and N5 variants, with this decrease being greater in voids than in sheets. These findings indicate that MG variants transfer matter more effectively from low density regions to high density regions. This results in the enhancement of density in knots, and simultaneously leads to emptier voids in the MG scenarios. This is also seen in the shift of the density peak in \cref{fig:distribution_all} to lower densities in MG when compared to \lcdm{}. Thus, more information can be gained via splitting by environment in the density distribution, rather than averaging across all environments. Since the MG variants enhance the higher densities in knots and further decrease lower densities in voids, the distribution in filaments happens to coincide with the transition region where the ratio is unity. This may change if we use a different threshold for the CW identification.

From this sub-section, we conclude that modifications to the underlying gravity theory impacts the distribution of density. Moreover, the density differences become more significant when studied individually across each CW environment. In the following section, we conduct a deeper investigation in the quantitative differences between these distributions by computing the higher-order moments of the density fields.

\subsection{Hierarchical clustering}
\label{subsec:cumulants}
The primordial Gaussian random field is symmetrical around the mean density, and positive and negative deviations from the mean density are equally probable. Studies of the mean and the two-point statistics of the density field (Correlation function, $\xi(r)$ or its Fourier transform, which is the power spectrum \pk{}) completely characterise such a Gaussian distribution. However, as the density field evolves, it becomes non-linear and highly asymmetric; positive density departures from the mean density can be very strong, while the negative deviations are restricted by the condition that the density, $\rho$, cannot be negative and hence the density contrast must obey the condition $\delta \geq -1$. This non-linear gravitational evolution yields a non-Gaussian distribution of the density at later times. As a result, relying solely on the two-point statistics is no longer enough to fully quantify the density distribution, as higher-order moments become significant \citep{lss_bernardeau,skewness_kurtosis_cosmic_density,Gaztanaga_1994,hivon_1995,higher_order_sdss,pdf_mg_de_ldt,3pcf_boss_dr12}.

In this context, we define the reduced moments, or the cumulants of the density distribution, $\langle \delta^i \rangle_c$. The cumulants are defined in terms of the central moments of the density distribution $\langle \delta^i \rangle$, and for the first four cumulants, we have \cite{smoothing_bernardeau}
\begin{equation}
\centering
    \begin{split}
        \langle \delta \rangle_{c} = \langle \delta \rangle = 0  \text{(mean)}, \\
        \langle \delta^2 \rangle_{c} = \langle \delta^2 \rangle \equiv \sigma^2  \text{(variance)}, \\
        \langle \delta^3 \rangle_c = \langle \delta^3 \rangle \equiv \mu_3 \text{(skewness)}, \\
        \langle \delta^4 \rangle_c = \langle \delta^4 \rangle - 3 \langle \delta^2 \rangle^2_c \equiv \mu_4 \text{(kurtosis)}.
       \end{split}
\label{eqn:cumulants}
\end{equation}
Here, $\langle \delta^i \rangle$ is given as
\begin{equation}
\label{eqn:moments}
    \langle \delta^i \rangle = \frac{1}{N}\sum_{n=1}^{N}(\delta-\langle\delta\rangle)^{i}.
\end{equation}

In this work, we focus on the first four cumulants of the density field distribution. Measuring the higher-orders becomes increasingly challenging due to the propagation of errors from the lower statistics,  which leads to increasing levels of uncertainty \cite{MG_SIGNATURES_HIERARCHICAL_CLUSTERING}.

In a Gaussian random field, all the cumulants, except for the variance, are zero. For a general random distribution, the first two non-vanishing cumulants after variance measure particular shape departures from a Gaussian distribution. For instance, $\mu_3$ (\textit{skewness}) quantifies the asymmetry of the distribution, and $\mu_4$ (\textit{kurtosis}) characterises the flattening of the tails of the distribution $w.r.t.$ a Gaussian. Higher-order moments measure more complicated shape deviations.

Various studies have revealed that for the case of initial adiabatic Gaussian density fluctuations, the gravitational instability produces a quasi-Gaussian clustering hierarchy of the cumulants of the density distribution \cite{lss_bernardeau}. Furthermore, the gravitational evolution of the initial Gaussian density field preserves this quasi-Gaussian clustering hierarchy, which is characterised by the \textit{hierarchical scaling} relations, given by

\begin{equation}
\label{eqn:reduced_cumulants}
    \langle \delta^n \rangle_c = S_n \langle \delta^2 \rangle_c^{n-1} = S_n\sigma^{2n-2}.
\end{equation}
Here, $S_n$ are the hierarchical amplitudes, or the reduced cumulants which are the moments factorised by the variance and higher powers. The first non-trivial reduced cumulant is of the third order \ie{} the reduced skewness, $S_3$, because $S_2 = 1$ when $n=2$. 

In both $f(R)$ and nDGP, the power spectrum is modified by the \textit{\textit{fifth-force}} dynamics, especially on the non-linear scales (\cite{pkmg_sg} and references within). Also, \cite{cw_pk_1,cw_pk_2} have shown that this two-point clustering statistic is impacted by the cosmic environment. Thus, we expect that the impact of MG on the two-point clustering statistic will be reflected in the higher-order moments, and also depend explicitly on the CW environment. For the case of our MG models that incorporate different screening mechanisms, it has also been shown that their modified dynamics in most cases leaves strong imprints on the matter clustering hierarchy, especially in the higher-order moments \citep{hierarchical_rebel,hierarchical_fR,MG_SIGNATURES_HIERARCHICAL_CLUSTERING,drozda_2022}. This work extends on the previous studies by additionally investigating the impact on different CW environments. 
 
In the next sub-sections, we present the results of our clustering studies on the density field with respect to different CW environments. As previously mentioned, we specifically focus on the $\sigma^2(R)$ (variance), $S_3(R)$ (reduced skewness), and $S_4(R)$ (reduced kurtosis). To  achieve these, we initially apply a top-hat smoothing filter to the density fields at different smoothing radii, $R$. Subsequently, we compute the clustering statistics for each CW element at every $R$.

\subsubsection{Variance}
\label{subsubsec:variance}

\begin{figure*}
    \centering
    \includegraphics[width=0.52\textwidth]{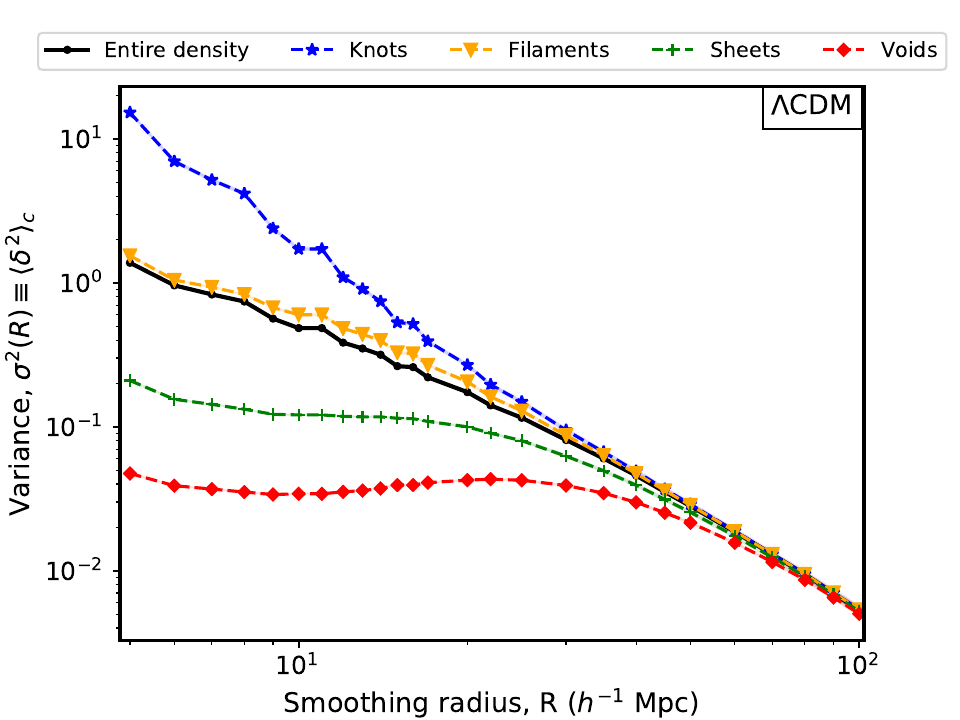}
    \includegraphics[width=0.495\textwidth]{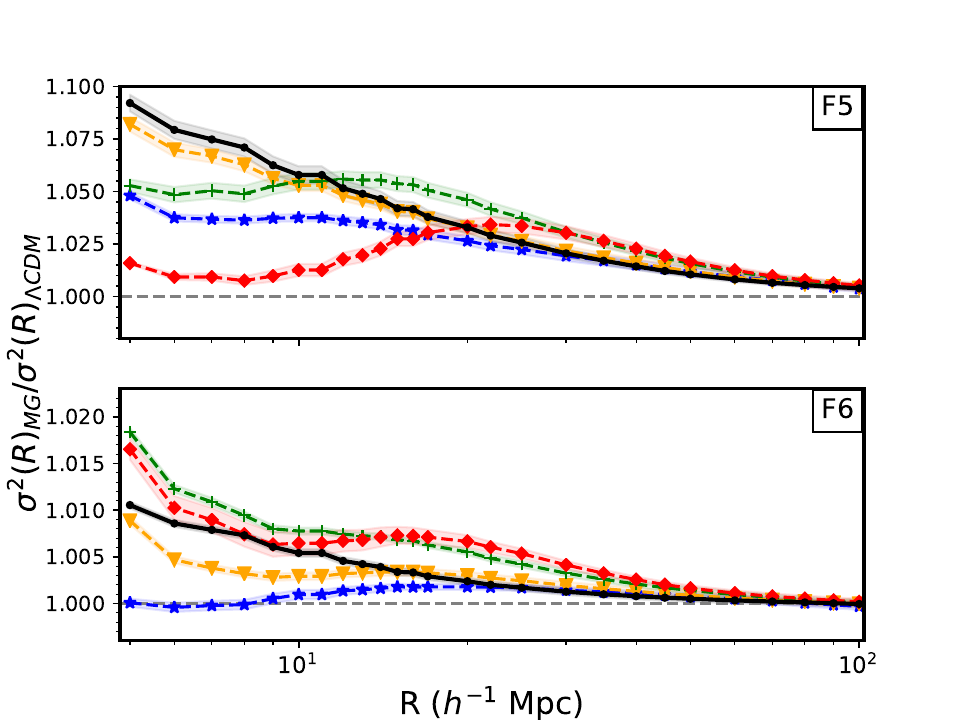}
    \includegraphics[width=0.495\textwidth]{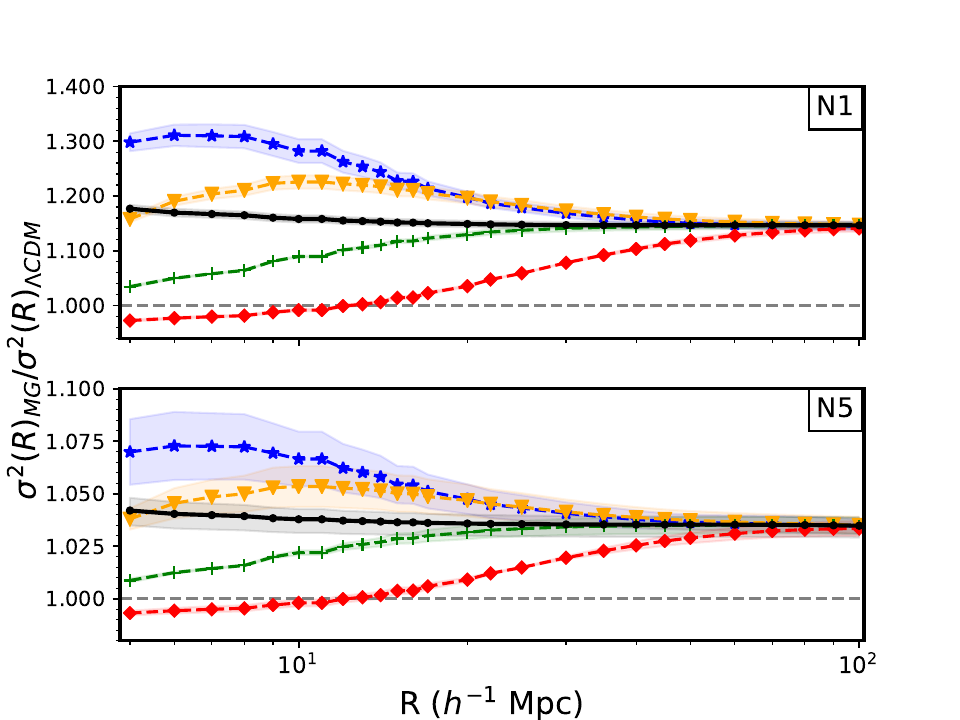}
\caption{\textit{Top plot:} Variance, $\sigma^{2}(R)$ for \lcdm{} across smoothing radii, $R$ ($h^{-1}$Mpc). The solid black line corresponds to the entire density field, and the coloured dashed lines are different CW environments. \textit{Bottom plots:} Comparison between MG and \lcdm{} variance, given by the ratio $\sigma^{2}(R)_{\text{MG}}/\sigma^{2}(R)_{\Lambda\text{CDM}}$ for the entire density and in different CW environment for the $f(R)$ (left) and nDGP (right) variants. Here, the shaded regions correspond to the standard deviation across five realizations.}
    \label{fig:variance}
\end{figure*}

In this sub-section, we show the second-order clustering characteristic, defined as the second cumulant in \cref{eqn:cumulants}, $\sigma^{2}(R)$ (\textit{variance}), as a function of smoothing scale, $R$, across CW environments. The results for variance are shown in \cref{fig:variance}. The top plot in this figure illustrates the behaviour of \lcdm{} variance in different CW environments. On smaller non-linear scales, we observe that knots exhibit the highest variance, followed by filaments (which closely aligns with the overall density distribution in black), and then sheets and voids. The variance in all these elements eventually converge towards the overall black trend at larger smoothing scales. This shows that the variance statistic is strongly dependent on both the environment and scale.

%overall density ratio
Similar trends are also observed in the corresponding plots for the MG models. In the bottom panels of \cref{fig:variance}, we plot the ratio $\sigma^{2}(R)_{\text{MG}}/\sigma^{2}(R)_{\Lambda\text{CDM}}$ across $R$. First, we discuss the black line which represents this ratio for the entire density field. Notably, both $f(R)$ and nDGP have an enhanced variance w.r.t. \lcdm{}, but show different trends. For the $f(R)$ models, we observe a scale-dependent trend in this ratio: a monotonic decrease as a function of $R$, which converges to unity on larger scales. In contrast, the overall density in nDGP model shows a systematic, scale-independent shift which is approximately 19\% for N1 and 4\% for N5. 

%for specific CW environments
Within both the $f(R)$ variants, each specific CW environment reveals a unique trend in the ratio. This suggests a complex, and a non-linear interplay between the scale-dependent \textit{fifth-force} and the environment-dependent chameleon screening mechanism. On the other hand, N1 and N5 ratios exhibit similar trends, albeit with different amplitudes. In these nDGP variants, knots, filaments, and sheets show an enhancement. On the other hand, voids exhibit a decrease in the ratio of the variance at small scales.

\subsubsection{Reduced Skewness, $S_3$}
\label{subsubsec:skewness}
\begin{figure*}
    \centering
    \includegraphics[width=0.52\textwidth]{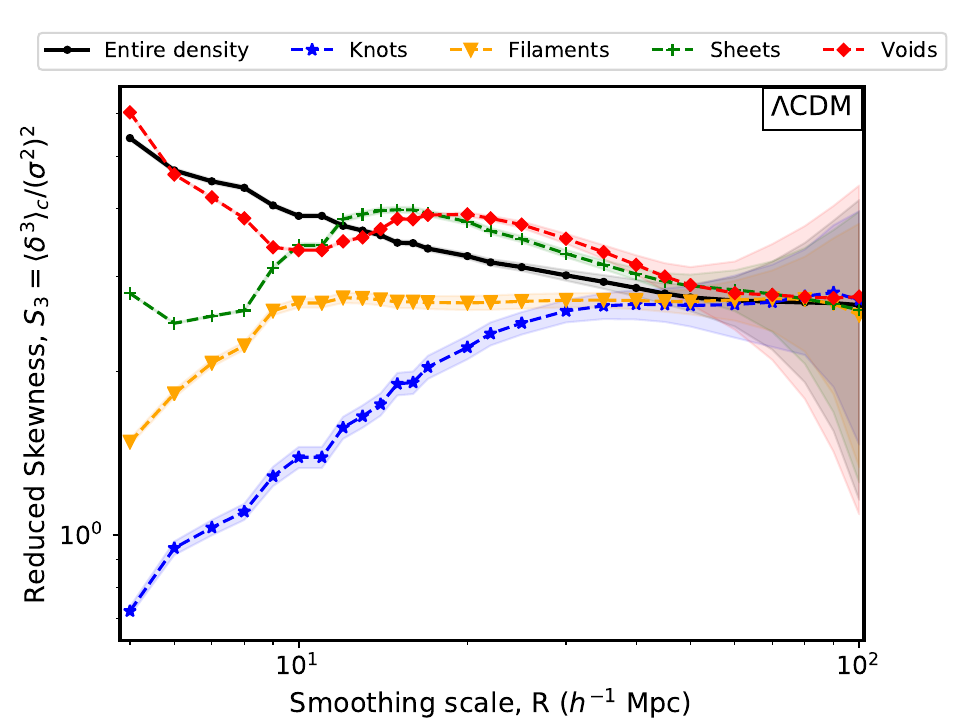}
    \includegraphics[width=0.495\textwidth]{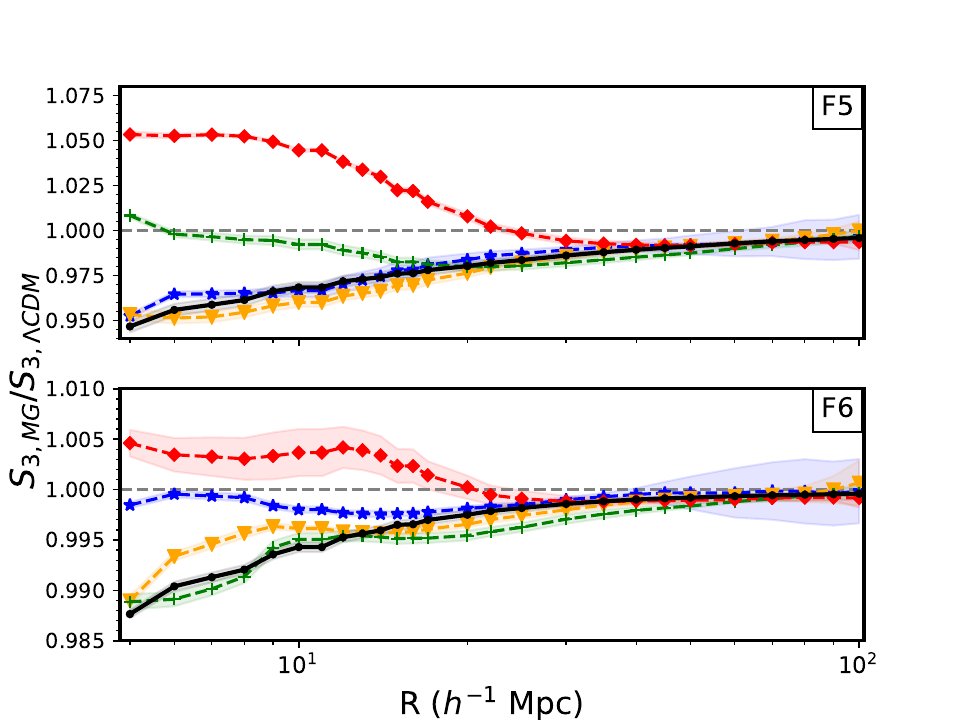}
    \includegraphics[width=0.495\textwidth]{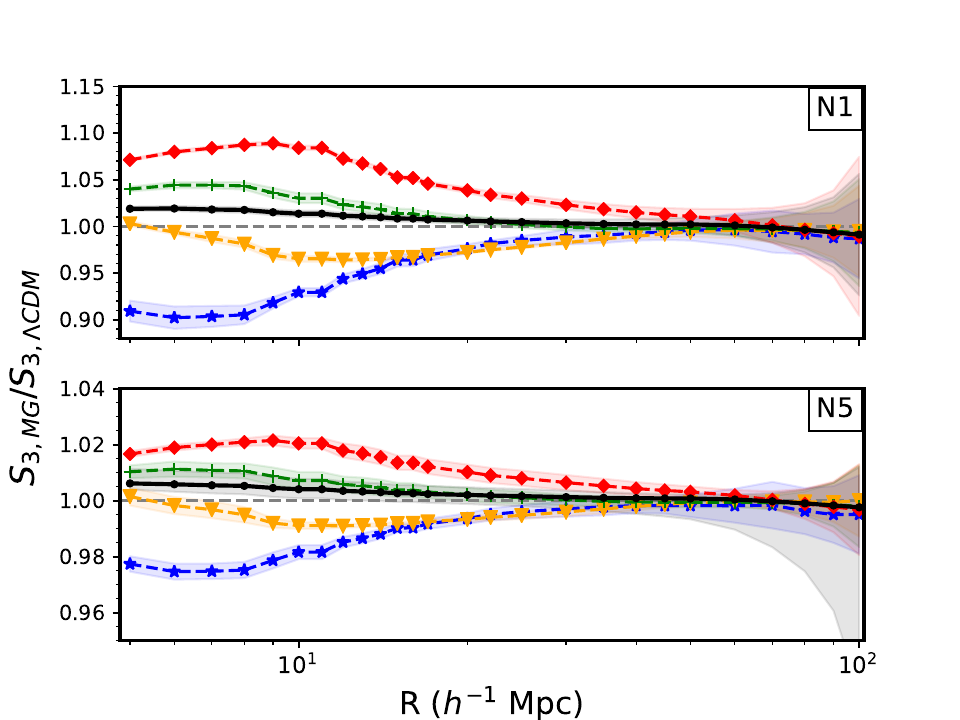}
\caption{\textit{Top plot:} Reduced skewness, $S_{3}$ for \lcdm{} across smoothing radii, $R$ ($h^{-1}$Mpc). The solid black line corresponds to the entire density field, and the coloured dashed lines are different CW environments. \textit{Bottom plots:} Comparison between MG and \lcdm{} $S_3$, given by the ratio $S_{3,\text{MG}}/S_{3,\Lambda\text{CDM}}$ for the entire density and in different CW environments for the $f(R)$ (left) and nDGP (right) variants. Here, the shaded regions correspond to the standard deviation across five realizations.}

    \label{fig:skewness}
\end{figure*}

In this and the following sub-section, we study the behaviour of reduced cumulants, $S_n$, (\cref{eqn:reduced_cumulants}). As previously mentioned, the first non-trivial reduced cumulant is of the third-order, which is referred to as the reduced skewness, $S_3$. In \cref{fig:skewness}, we show similar results as \cref{fig:variance}, but for $S_3$. Namely, in the top plot, we show the $S_3$ for \lcdm{} across different CW environments. Here, again we have a clear environmental dependence, with knots showing the smallest value of $S_3$ on small scales. This opposite result to the variance plot can be attributed to the factor of variance and its orders in the denominator of computing $S_n$ (\cref{eqn:reduced_cumulants}). Voids here have enhanced values of $S_3$, and are closer to the overall density trend. As we move to the larger scales, the value of $S_3$ converges for all environments. The increase in uncertainty is expected when we compute higher cumulants as the error from lower moments propagate in the higher-order statistics. Furthermore, the uncertainty also increases at larger scales due to cosmic variance.

In the bottom two plots of \cref{fig:skewness}, we plot the ratio $S_{3,\text{MG}}/S_{3,\Lambda\text{CDM}}$ for $f(R)$  and the nDGP variants. In all the MG models, there is a clear environmental dependence on small scales, and the ratios approaches unity as the smoothing scales increase. Voids have the maximum enhancement in all models. For the high density knots and filaments, the reduced skewness decreases in all MG models when compared to \lcdm{}. Additionally, sheets and the overall density in $f(R)$ also have a decreased value of $S_3$ compared to \lcdm{}. Similar to the variance plot, $f(R)$ shows a more complicated behaviour across the two models. For the case of nDGP, the ratio across all environments is again similar in both models, however with different amplitudes. For this MG model, voids and sheets have enhanced $S_3$, and knots and filaments have a decrease in $S_3$ across the considered smoothing scales.

\subsubsection{Reduced Kurtosis, $S_4$}
\label{subsubsec:kurtosis}
\begin{figure*}
    \centering
    \includegraphics[width=0.52\textwidth]{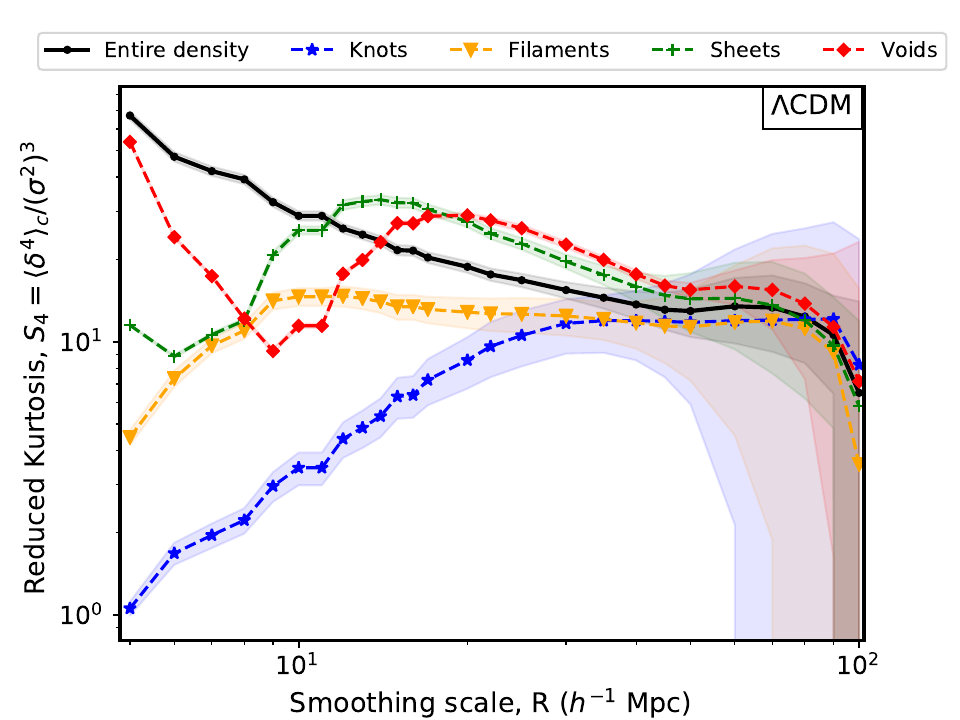}
    \includegraphics[width=0.495\textwidth]{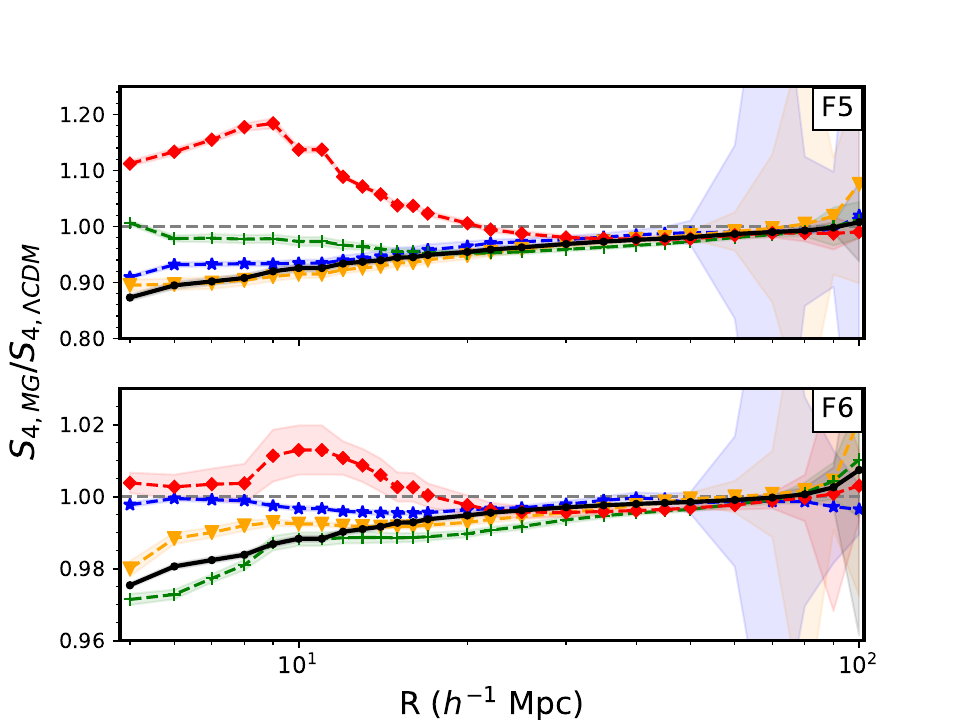}
    \includegraphics[width=0.495\textwidth]{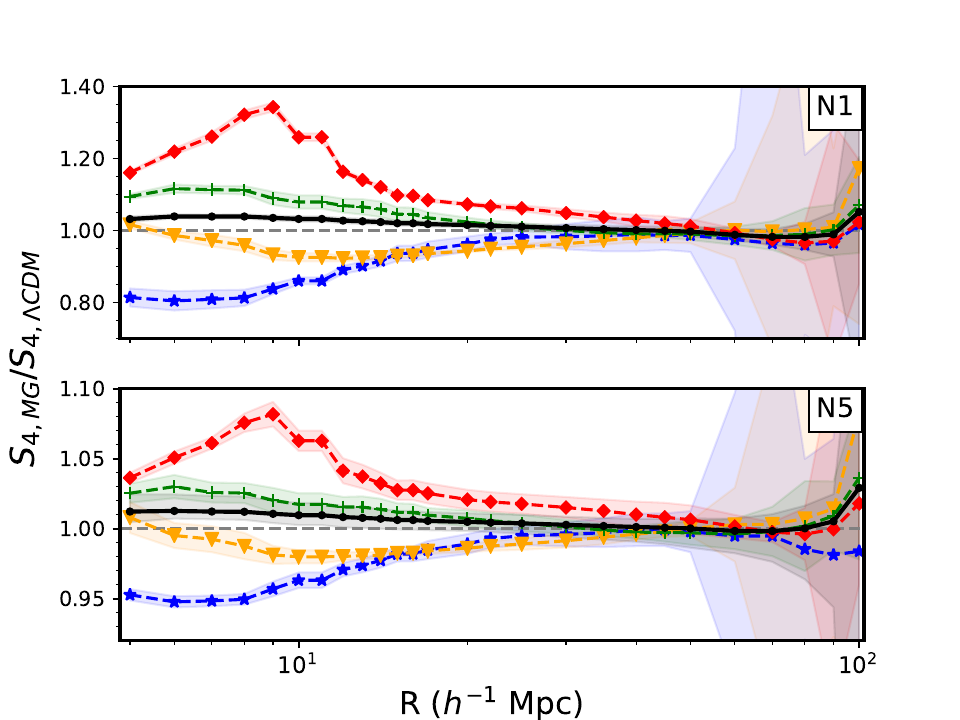}
\caption{Same as \cref{fig:skewness}, but for $S_4$. 
}
    \label{fig:kurtosis}
\end{figure*}

We extend our analysis by further computing the fourth reduced cumulant ($n=4$ in \cref{eqn:cumulants}), which is the reduced kurtosis, $S_4$. The results for our $S_4$ computations are shown in \cref{fig:kurtosis}. In the top plot of this figure, we again have the \lcdm{} $S_4$ across different CW environments. The environment-dependent trend in the $S_4$ value across $R$ is similar to the corresponding plot of $S_3$ in \cref{fig:skewness}, however with enhanced features as well as larger uncertainties. 

In the bottom plots of the same figure we see that the impact of MG on $S_4$ is more pronounced when compare to the $S_3$ statistics. Also, nDGP has a greater impact on $S_4$ than $f(R)$. The relative effect, similar to the previous case of $S_3$, shows the strongest environmental dependence on small scales with decrease in MG $S_4$ in knots, filaments and sheets for $f(R)$, and in knots and filaments for nDGP. Ratios for all environments converge to unity as we go to larger scales. $S_4$ in voids shows the maximum enhancement in all MG models, with a peak-like feature at $\approx 10$ \Mpch{}. It remains uncertain whether this observed trend is an artefact of the numerical simulations or is a physically meaningful signal. A deeper investigation of this trend requires higher resolved simulations.

The \textit{fifth-force} dynamics, while enhancing the density variance $w.r.t.$ to the fiducial \lcdm{} case (\cref{subsubsec:variance}), actually leads to lower skewness and kurtosis (and higher-order cumulants as well, as shown previously in \citep{hierarchical_rebel,hierarchical_fR,MG_SIGNATURES_HIERARCHICAL_CLUSTERING,drozda_2022}). There is an intrinsic asymmetry imprinted in the density field \ie{} on the positive side of the distribution, the density can grow to arbitrarily large values, reaching presently up to orders of $\approx 10^6$ in the centres of cluster-sized DM halos. However, \cref{fig:mean_distribution_cw} shows that voids in MG models tend to be much emptier than in \lcdm{} scenarios, which results in the shift of the PDF of density in MG to lower values, as seen in \cref{fig:distribution_all}. This is expected to lead to an overall decrease in the reduced asymmtery between the tails of the density, and in turn in the skewness and kurtosis for MG $w.r.t.$ \lcdm{}. This trend is seen in the $f(R)$ variants (left plots of \cref{fig:skewness,fig:kurtosis}). However, for nDGP models, we see an overall increase in these reduced cumulants, particularly at smaller length scales (corresponding right plots of these figures). This variation can be explained by studying these statistics separately in different environments, as indicated by the coloured lines in these figures. In the $f(R)$ models, all CW environments, except voids, exhibit a decrease in $S_n$ compared to \lcdm{}. In contrast, in the nDGP models, both voids and sheets show enhanced values for these statistics, while knots and filaments show a decrease. This additional information from each environment separately provides a deeper insight into the differences in the trends observed across MG models, which have different underlying \textit{fifth-force} physics and screening mechanisms.

Perturbation theory for GR shows that the reduced cumulants are a decreasing function of the smoothing scales, reflecting that the density fields become more Gaussian as the fluctuations are smoothed out \citep{lss_bernardeau}. We find these results for both \lcdm{} and MG models, as well as for all the CW environments. Also, the reduced cumulants have shown to be sensitive to the spectral index of the initial power spectrum \cite{smoothing_bernardeau}. This makes the studies of higher-order clustering a probe to constrain the initial density distribution \cite{juszkiewicz_1993}. Furthermore, our results from analysing the higher-order statistics show that at the relevant scales for galaxy and halo formation, both the $f(R)$ and nDGP density fields are characterised by change in the clustering across the correlation orders that we studied, and this change depends explicitly on the CW environment. Thus, the CW environment offers a significant systematic, and hence should be accurately catered to in order to make unbiased estimates for measures that are associated with the studies involving growth rate of structure formation, primordial non-Gaussianity, galaxy formation models, and large-scale clustering \citep{lss_bernardeau,juszkiewicz_skewness_bao,wl_mg_Schmidt_2008,PNG_MOMENTS,png_moments_2,cw_pk_1,cw_pk_2}.

\subsection{Halo mass function}

In the preceding sub-sections, we explore the influence of the \textit{fifth-force} on the density distribution, and on the clustering statistics of underlying density fields in different CW environments. In this and the subsequent sub-sections, we show the impact of the \textit{fifth-force} on halo properties across CW environments. Many past works have separately shown significant influence of both the \textit{fifth-force} \citep{Schmidt_2009_fR,Schmidt_Oyaizu_2008_3,borderline_f6_hmf,mitchell_fR,mitchell_ndgp,hmf_sg,spin_mg_2,mg_spin_fR}, and the CW \citep{,hellwing_CW,hahn_cw_2007,conc_spin_shape_CW,evolution_cw_nexus,abundance_cw_2,halo_abundance_cw} on halo properties. Here, we further describe how the statistics associated with halos are impacted by the combined impact of the MG and different CW environments. For this purpose, we segment the entire halo population into knots, filaments, sheets, and voids, depending on their spatial location. Here, we probe the property of halo mass function, and in the next sub-section we discuss the impact on halo spin. To compute these halo properties, we categorise the halo masses in specific mass bins. We also test the effect of varying the binning on both the HMF and spin, and find negligible influence of varying the mass bins on these halo properties.

\label{subsec:hmf}
\begin{figure*}
    \centering
    \includegraphics[width=0.52\textwidth]{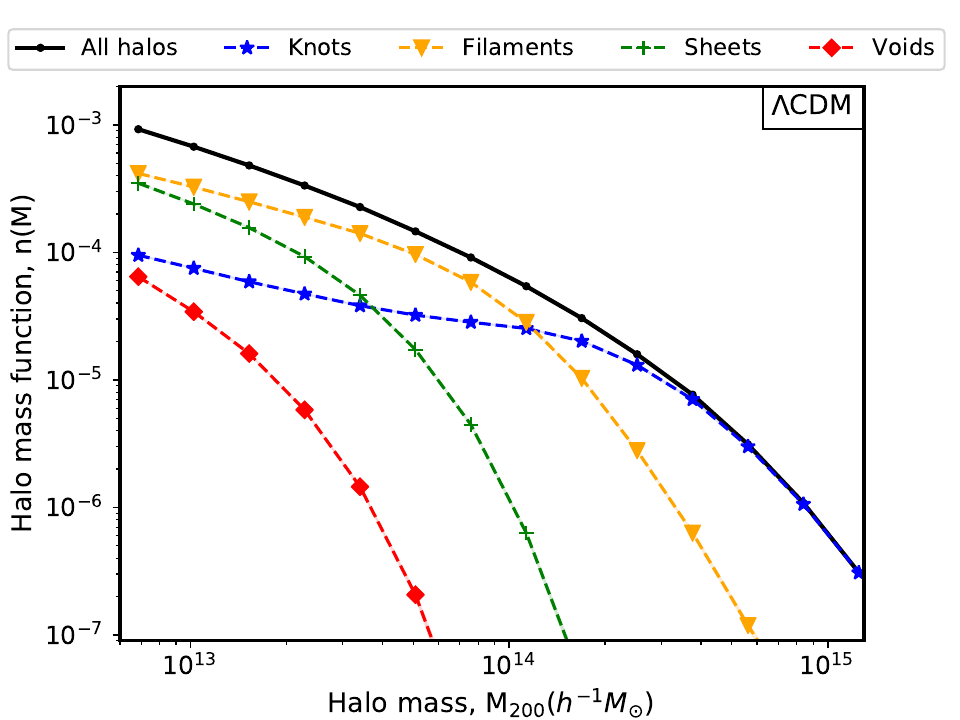}
  \includegraphics[width=0.495\textwidth]{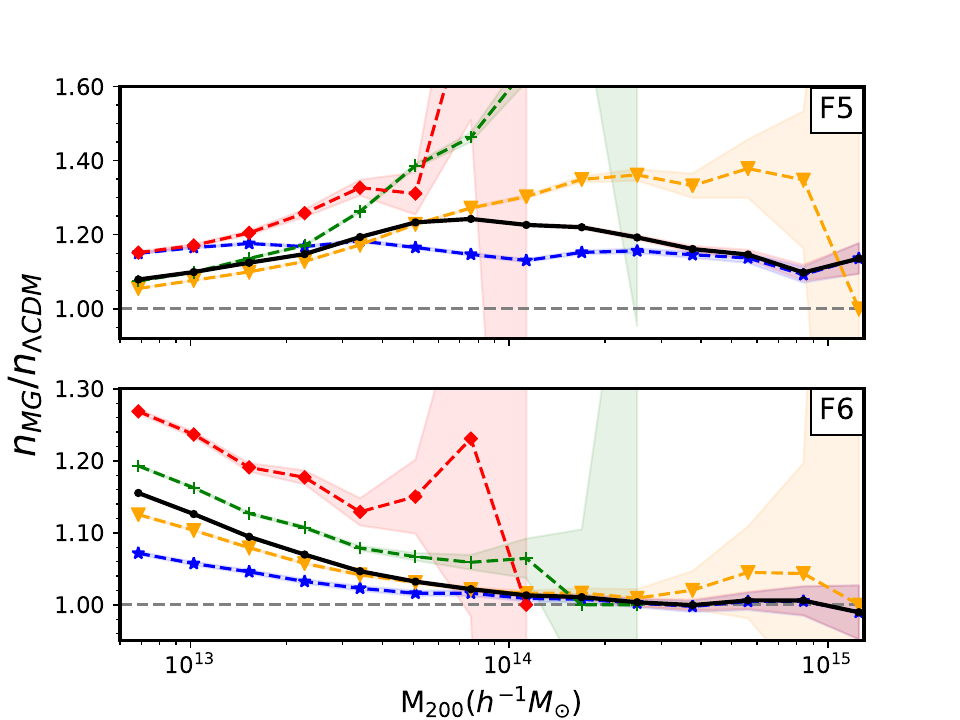}
  \includegraphics[width=0.495\textwidth]{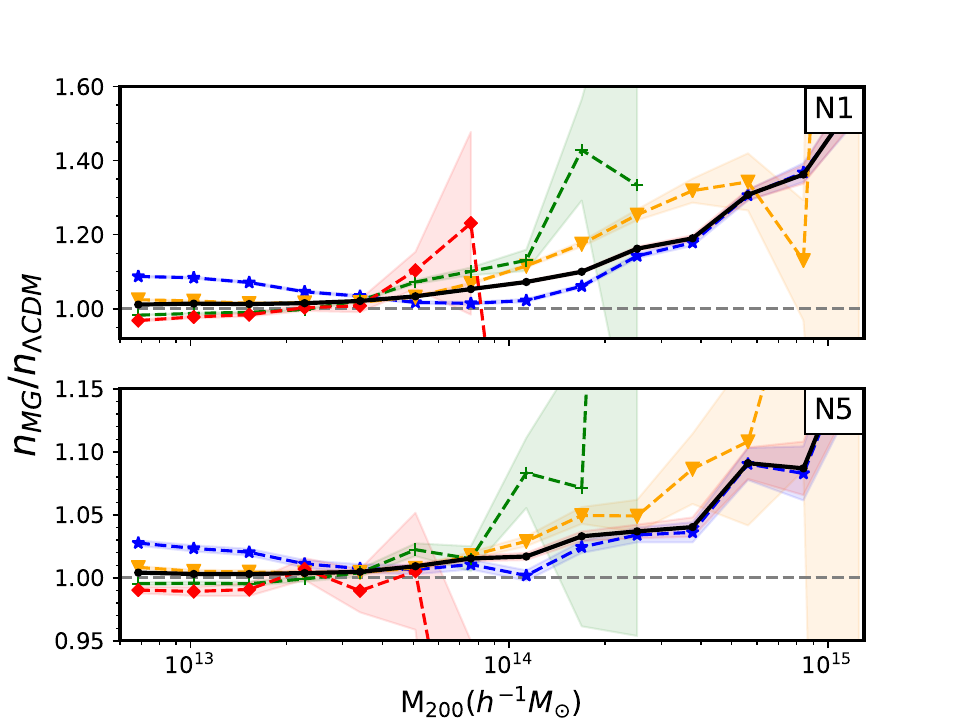}

\caption{\textit{Top plot:} Halo mass function (HMF) (\cref{eqn:HMF_MAIN}), for \lcdm{}, expressed as a function of halo mass, $M_{200}$ (\msunh{}). Here, the solid black line represents the HMF obtained for the entire halo populations, and the dashed lines of different colours represent halos separately in each CW environment. The shaded regions are the Poisson errors obtained from the halo number count in each mass bin. \textit{Bottom plots:} Ratio of the HMF between different MG models to \lcdm{} across halo masses, for the entire halo populations (black lines) and for halos in different CW environments (coloured dashed lines). The left plots are the ratio for the $f(R)$ variants, and the right plots are the nDGP ratio results. Here, the shaded regions illustrate propagated Poisson errors obtained from the halo number count.} %\PGVc{Better to keep the y-labels consistent throughout- to either have one digit after decimal or two.}}
    \label{fig:hmf_all}
\end{figure*}

The study of the halo mass function (HMF) plays a crucial role in understanding and validating the current theory of LSS and galaxy formation. It is by now clear that the efficiency of halo formation depends on the environment in which the halo resides \citep{hahn_cw_2007,halo_abundance_cw,vweb_yehuda,evolution_cw_nexus,hmf_fR_cw,tracing_cw_libeskind_2018,hellwing_CW}. We begin by studying the influence of CW environment on halos by plotting the HMF segregated into different web environments in \cref{fig:hmf_all}. The HMF is defined as the number density of halos in a given mass interval ($d \log M$) as given in the following equation,
% \begin{equation}
% \label{eqn:HMF_MAIN}
% n(M) \equiv  \frac{dn}{dM} = \frac{\rho_{m}}{M^2} F(\sigma) \left|\frac{d\ln \sigma}{d\ln M} \right|.
% \end{equation}

\begin{equation}
\label{eqn:HMF_MAIN}
n(M) =  \frac{1}{V}\frac{dN}{d\log M}.
\end{equation}

Here, $dN$ corresponds to the number of halos in each mass bin, and $V$ is the volume of the simulation.

In the top plot of \cref{fig:hmf_all}, we show the HMF for \lcdm{} across different CW environments. The black lines correspond to the HMF obtained for the entire halo population and different colours depict the halo abundance in each CW element. This plot shows that the HMF exhibits a substantial difference between environments. 
Massive haloes with $M > 10^{14}$ \msunh{} reside in clusters or knots, while filaments host most of the halos between $7.8 \times 10^{12}-10^{14}$\msunh{}. Sheets and voids do not dominate the halo population in the mass ranges shown here, showing a sharp drop in abundance at consecutively lower masses compared to filaments. Similar results for \lcdm{} have also been reported in previous works (e.g. \cite{evolution_cw_nexus,tracing_cw_libeskind_2018,hellwing_CW,cosmic_ballet_1, cosmic_ballet_3}).

In both $f(R)$ and nDGP, the additional \textit{fifth-force} leads to enhanced structure formation. As a result, we have an increase in the MG HMF $w.r.t.$ standard \lcdm{} for the halo mass range examined in this work. We show the MG to \lcdm{} HMF in the lower plots of \cref{fig:hmf_all}. The left plots show the $f(R)$ gravity, and the right are the nDGP gravity results. Both these MG models illustrate different trends in the HMF ratio $w.r.t$ \lcdm{}. In all these plots, the solid black lines correspond to the ratio of MG to \lcdm{} HMF for the entire halo population. In F5, there is a peak-like feature in this ratio, whereas in F6 we see that this ratio monotonically decreases with halo mass. The peak-like feature in the intermediate mass scales for the $f(R)$ variants is a result of environment screening at low-mass halos and self-screening in the halos with higher masses \cite{hmf_sg}. Meanwhile, for both the nDGP models, we see that the ratio increases with halo mass. In this model, as the halo mass and size of the halo increases, more matter sits outside the Vainshtein radius of the halo. This would increase the impact of the \textit{fifth-force} on the halo, thereby increasing the abundance of higher mass halos. These HMF trends for the entire halo populations have been previously discussed in \cite{mitchell_fR,mitchell_ndgp,hmf_sg,sg_hmf_pta}.

Furthermore, the HMF ratio across each environment is represented as a dashed line of different colour, and the shaded regions indicate the propagated Poisson errors. In both the $f(R)$ models, voids have the maximum HMF enhancement at small halo mass scales. The low densities of voids result in reduced overall screening, thereby amplifying the effects of the\textit{ fifth-force.}

In F5, this enhancement in voids is followed by knots, then sheets and then filaments. Whereas, in F6, we see that the void enhancement is followed by sheets, then filaments, and then knots. For both these $f(R)$ variants, the ratio for the overall HMF (black line) lies between filaments and sheets for $M_{200} \leq 5 \times 10^{13}$\msunh{}, and follows the trend of knots for larger halo masses. This trend closely follows the absolute HMF plot. Also, it is evident here that for large halo masses, the HMF decreases $w.r.t$ \lcdm{} when considering all halos (black line). This can be attributed to self-screening, where the decrease in the HMF is more significant in F6 than F5, as the screening is more dominant over the effect of the \textit{fifth-force} in the weaker F6 variant.

Contrary to $f(R)$, voids and sheets in nDGP show a suppressed HMF $w.r.t.$ \lcdm{} for $M_{200} \leq 3 \times 10^{13}$ \msunh{}. With increase in halo mass, the size of halos also increases. As a result, larger mass halos, which mostly reside in knots and filaments, would be larger than the Vainshtein radius and have a reduced impact from the Vainshtein screening, and in-turn would experience the \textit{fifth-force} more strongly. Hence, the HMF for voids and sheets in nDGP (which mostly host low-mass halos) is least enhanced. Here, similar to $f(R)$, the overall HMF trend (black line) lies between filaments and sheets for $M_{200} \leq 5 \times 10^{13}$\msunh{}, and follows the trend of knots as the halo mass increases. 

The nonlinear trends in these bottom four plots convey important insights, emphasizing that the \textit{fifth-force} exerts distinct impacts on the HMF across different CW environments, and at different halo mass scales. Incorporating the environmental effects can provide a more powerful method to disentangle the complex relationship between MG models and the cosmic environments hosting halos.

 \subsection{Halo angular momentum} 
\label{subsec:spin}

The angular momentum, or spin, of a halo is defined as the sum of the angular momentum of all individual particles that make up the halo, \ie{}
\begin{equation}
  \textbf{J} = \sum_{i=1}^{N} m_i (r_i \times v_i),  
\end{equation}
where $m_i$ is the particle mass, $r_i$ and $v_i$ are the position and velocities of the $i^{th}$ particle $w.r.t$. the centre of mass of the halo. 
The amplitude of the angular momentum, $J = \left| \textbf{J} \right|$ is usually expressed in terms of a dimensionless spin parameter, $\lambda_p$, introduced by \cite{peebles_1969}, and is given as
\begin{equation}
\label{eqn:spin}
    \lambda_p = \frac{J|E|^{1/2}}{GM_{200}^{5/2}}.
\end{equation}

The $\lambda_p$ parameter characterises the overall importance of angular momentum relative to random motions. Halos with low values of $\lambda_p$ are dominated by velocity dispersion due to random motions, and are said to have low spin. On the other hand, halos with higher $\lambda_p$ rotate faster and show coherent orbital rotations and are therefore said to have high spin. 

In this work, we use the definition of halo spin given by Bullock et al. (2001) \cite{spin_bullock_2001}. This expression offers greater practicality as it removes the necessity to explicitly compute the energy of the halo in \cref{eqn:spin}. In this case, for a mass $M_{200}$ enclosed within a region $R_{200}$, $\lambda$ is given by
\begin{equation}
\label{eqn:spin_bullock}
    \lambda = \frac{J}{\sqrt{2}M_{200}V_{200}R_{200}}.
\end{equation}
Here, $J( = \sqrt{J_x^2+J_y^2+J_z^2})$ is the angular momentum within $R_{200}$, which encodes the MG effect. $\lambda$ reduces to $\lambda_p$ at the virial radius of a truncated isothermal halo. All these halo properties can be obtained directly from the \textsc{rockstar} halo catalog. 

Halo angular momentum is a result of non-vanishing torque generated by the gravitational field of the surrounding matter distribution in the earlier phases of halo formation (termed as the protohalo). This highlights the dependence of spin on the underlying gravitational theory, and hence this property can be an independent discriminator for testing MG signatures. Previous studies 
(\citep{spin_ide, hmf_fR_cw, mg_spin_fR, spin_mg_2, rebel_hellwing} to cite a few) have shown that this is indeed true, and spin is sensitive to modifications to the standard gravity theory, and increases with enhancement in gravity. This is expected as a greater gravitational force would result in a greater torque, and in the enhancement of the overall spin of a halo.

\subsubsection{Probability Distribution of spin}
\label{subsubsec:pdf_spin}

\begin{figure}
    \centering
    \includegraphics[width=\columnwidth]{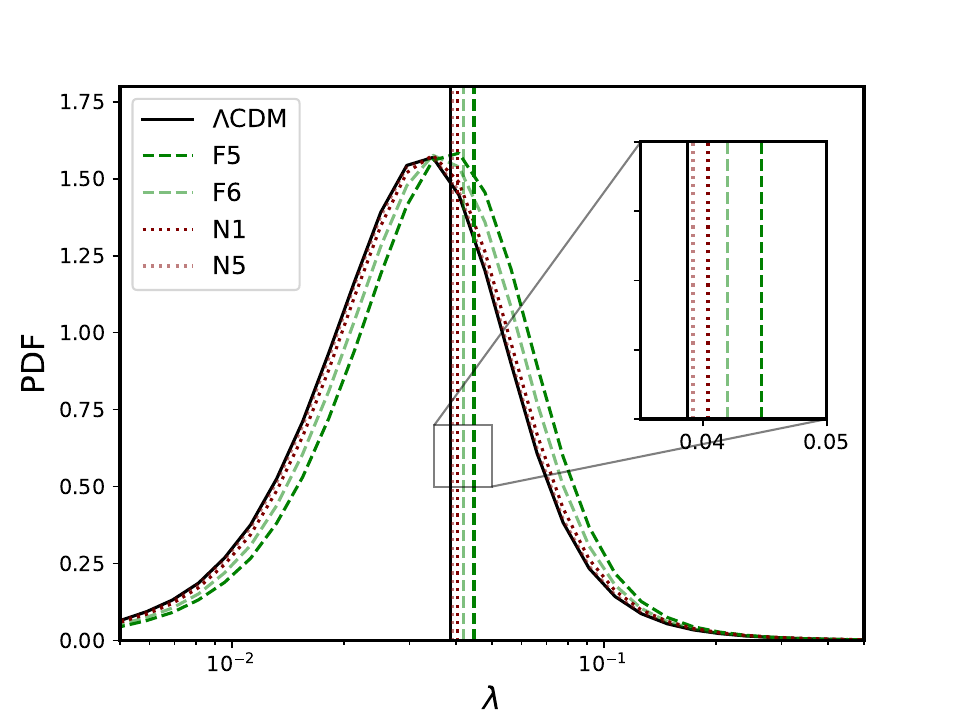}

    \caption{ %\PGVc{Same suggestion as for figure 1 wrt to the colors and linestyles. I would also make the lines in the inlet plot thicker.}
    PDF for the spin parameter, $\lambda$ defined in \cref{eqn:spin_bullock} for all the cosmologies considered in this work. The mass range for this distribution is $5 \times 10^{13} - 2 \times 10^{14}$ \msunh{}. The vertical lines are the mean values of the $\lambda$ for each cosmology, with the inset showing the zoomed version of the mean values.}
    \label{fig:distribution_all_spin}
\end{figure}

To probe the quantitative differences between spin in MG and \lcdm{}, we plot the distribution of the spin parameter $\lambda$ for all the cosmologies in \cref{fig:distribution_all_spin}. The mass range of the halos considered  to calculate this distribution is $5 \times 10^{13} - 2 \times 10^{14}$ \msunh{}, as they have $\geq 300$ particles, which is a reliable convergence limit for spin computations \cite{300particles_spin}. The vertical lines in this plot are the mean values of $\lambda$ for each model. The inset axes illustrates a clear difference in the mean values for all the cosmologies. The mean spin is within 0.035 - 0.045, with the MG values enhanced when compared to \lcdm{}. Stronger variants (F5 and N1) exhibit greater difference from \lcdm{} compared to their respective weaker counterparts (F6 and N5).

To further check the statistical significance of this spin enhancement in MG, we use the property that the spin parameter distribution is well approximated by a log-normal distribution, given by \cite{barnes_efstathiou_1987,spin_bullock_2001}

\begin{equation}
    \label{eqn:lognormal_spin_eqn}
    p(\lambda) d\lambda = \frac{1}{\lambda\sigma\sqrt{2\pi}}\text{exp} \left[ -\frac{(\text{ln}\lambda-\mu)^2}{2\sigma^2} \right] d\lambda.
\end{equation}

Here, the mean value of the log-normal distribution is given by
\begin{equation}
    \lambda_0 = e^{\mu + \frac{1}{2}\sigma^2},
\end{equation}

and the standard distribution is
\begin{equation}
    \sigma_{\lambda} = e^{\mu + \frac{1}{2}\sigma^2}\sqrt{e^{\sigma^2}-1}.
\end{equation}

In \cref{tab:spin_fit_params}, we show the best-fit values of the parameters $(\lambda_0, \sigma_\lambda)$ for the log-normal distribution of the $\lambda$ parameter in all the cosmologies under consideration. These best-fit parameters were obtained by solving the minimum reduced $\chi^2$, obtained by fitting the spin PDF from simulations with \cref{eqn:lognormal_spin_eqn}. This table clearly shows that both the mean, $\lambda_0$, and the standard deviation, $\sigma_\lambda$ increases as we move from \lcdm{} to the stronger MG models. 

\begin{table}
    \centering
    \begin{tabular}{||c|c|c||}
    \hline
    \hline
    Model & $\lambda_0$ & $\sigma_\lambda (\times 10^{-3})$ \\
    \hline
     \lcdm{}    &  0.037 & 0.220 \\
    \hline
     F5    & 0.043 & 0.300 \\
    \hline
    F6     & 0.040 & 0.260 \\
    \hline
     N1    & 0.038 & 0.235 \\
    \hline
     N5    & 0.037 & 0.223   \\
    \hline
    \hline
    \end{tabular}
    \caption{The mean, $\lambda_0$ and standard deviation, $\sigma_{\lambda}$ for the best log-normal fit to the halo spin distributions for each gravity model (\cref{eqn:lognormal_spin_eqn}). Here, we have considered halos in the mass range $5 \times 10^{13} - 2 \times 10^{14}$ \msunh{}.}
    \label{tab:spin_fit_params}
\end{table}

\subsubsection{Impact of MG on the evolution of halo spin}
\label{subsubsec:MG_spin}
In the previous sub-section, we demonstrate that the spin increases with an enhancement in gravity. Here, we show the physical explanation for this enhancement. According to the TTT framework, the $i^{th}$ component of the angular momentum tensor is
\begin{equation}
\label{eqn:spin_growthrate}
    J_i(t) = a^2(t)\dot{D}(t)\epsilon_{ijk}T_{jl}I_{lk}.
\end{equation}
Here, $I_{lk}$ is the inertia tensor of the protohalo. $T_{jl}$ is the Hessian, constructed from partial derivatives of the density of the protohalo, which makes it dependent on the neighbouring density distribution. This expression reflects that the angular momentum of a protohalo is the tensor product of the inertia tensor and the tidal tensor. Here, $a(t)$ is the scale factor, and $\dot{D}(t)$ corresponds to the change in the growth rate $D(t)$ with time. By maintaining fixed initial conditions across all MG models \ie{} fixed $T$ and $I$,
\begin{equation}
    J(t) \propto a^{2}(t) \dot{D}(t).
\end{equation}

\begin{figure*}
    \centering
    \includegraphics[width=\textwidth]{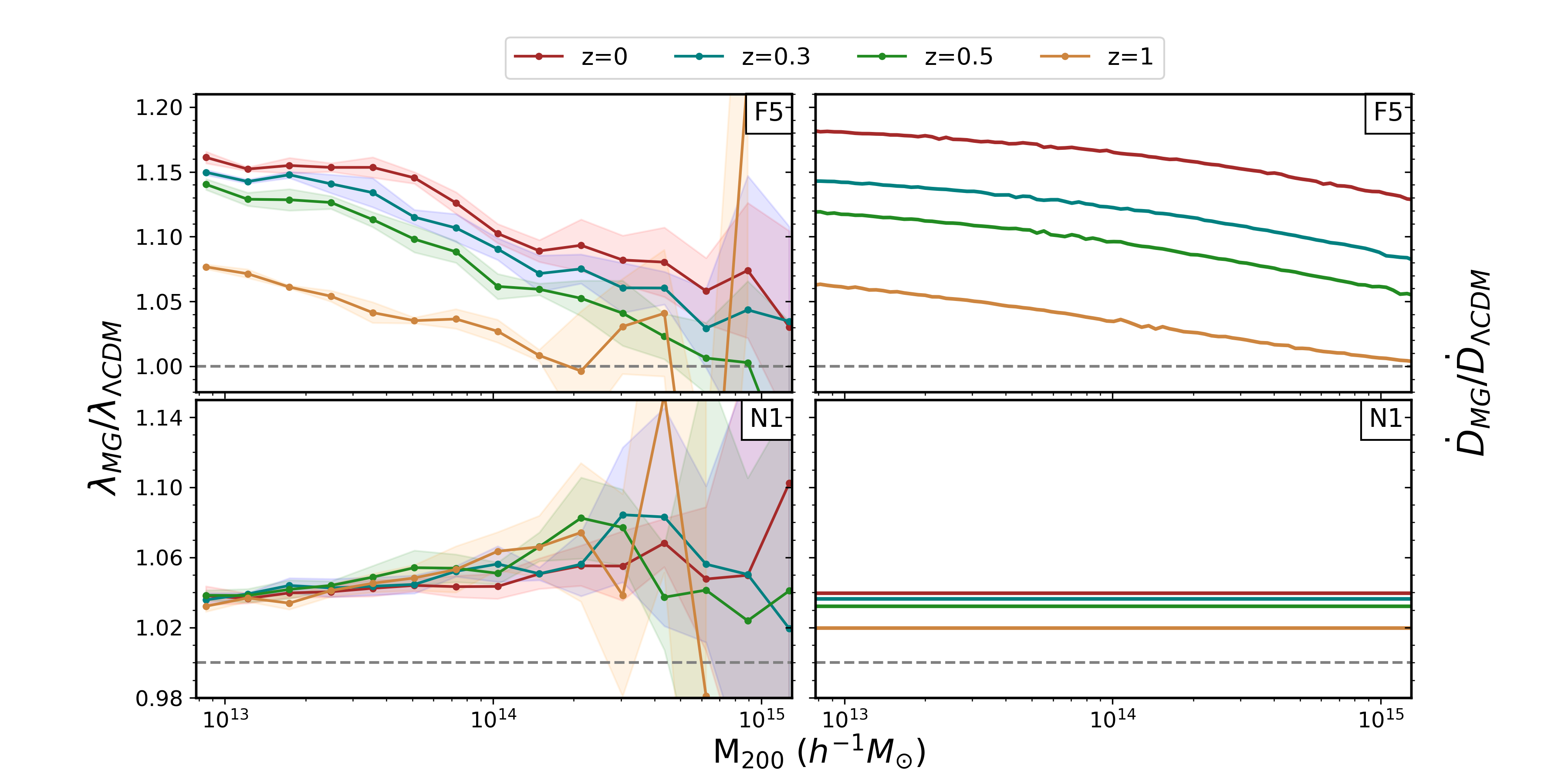}

\caption{\textit{Left plot:} Ratio of the halo spin, $\lambda$, between MG and \lcdm{}, as a function of halo mass, $M_{200}$ (\msunh{}), for different redshifts from simulations. \textit{Right plot:} Corresponding ratio of the rate of change of growth factor, $\dot{D}(t)$ between MG and \lcdm{}, as a function of $M_{200}$ (\msunh{}). The top panel in both plots corresponds to the F5 model, and the bottom panel is the N1 result.}

    \label{fig:spin_growthrate_MG}
\end{figure*}

This above expression illustrates a direct relation between the angular momentum of halos, $J(t)$, and the change in growth rate with time, $\dot{D}(t)$. Hence, TTT suggests that the MG effect on spin can be captured nicely by considering the growth factor, which gets a boost in both $f(R)$ and nDGP. To test this, we plot \cref{fig:spin_growthrate_MG}. In the left plot of this figure, we show the spin ratio from simulation, $\lambda_{\text{MG}}/\lambda_{\Lambda \text{CDM}}$ as a function of halo mass and redshift. In the top panel, we show the ratio $\lambda_{\text{F5}}/\lambda_{\Lambda \text{CDM}}$ as a function of halo mass and redshift, which decreases as both increase. In the case of N1 model in the bottom panel, we observe a small and a constant enhancement of $\lambda$, which is largely independent of both the halo mass and redshift. Here, we only show the stronger variants as the weaker models exhibit similar trends with reduced values. 
%\spcc{Need to explain how to calculate the growth factor as a function of mass as it is typically expressed as a function of wavelength, k.}

The right plot in this figure shows the ratio $\dot{D}_{\text{MG}}/\dot{D}_{\Lambda \text{CDM}}$ for the same redshifts. To calculate the growth rate $D(t)$ across redshifts, we use the input of linear power spectrum from modified gravity version of \textsc{camb} \cite{MGCAMB2011JCAP}, where the growth rate for a given redshift is the ratio of the square root of the power spectrum at that redshift to the present. This quantity is scale-independent in nDGP, and in the scale-dependent $f(R)$ model is a function of wavenumber. We compute the Lagrangian halo mass corresponding to this wavenumber and plot the ratio as a function of halo mass.

On comparing both the ratio plots in this figure for a given MG model, we observe a clear correlation between spin $\lambda(t)$ (or $J(t)$), and the rate of change in the growth enhancement, $\dot{D}(t)$. There is both quantitative and qualitative similarity between these ratios across the halo mass scales and redshifts. This similarity affirms a direct physical relation between the change in the linear growth factor in MG models $w.r.t.$ \lcdm{}, and its impact on halo spin. 

Using TTT, we can separate the initial tidal field and the cosmological evolution (\cref{eqn:spin_growthrate}). As the initial conditions are identical between the different cosmological models, \ie{} the initial tidal field is fixed, TTT predicts that the spin evolution is determined by the cosmological model. This appears to be generally correct, although we do not capture the full MG effect on spin by considering only the change in the growth factor. Additional work is required to comprehensively describe the remaining effects of the MG physics on the spin. This would pave way for the development of semi-analytical models to better understand this relationship.

\subsubsection{Impact of MG on halo spin in CW environments}
\label{subsubsec:spin_mg_cw}

\begin{figure*}
    \centering

    \includegraphics[width=0.52\textwidth]{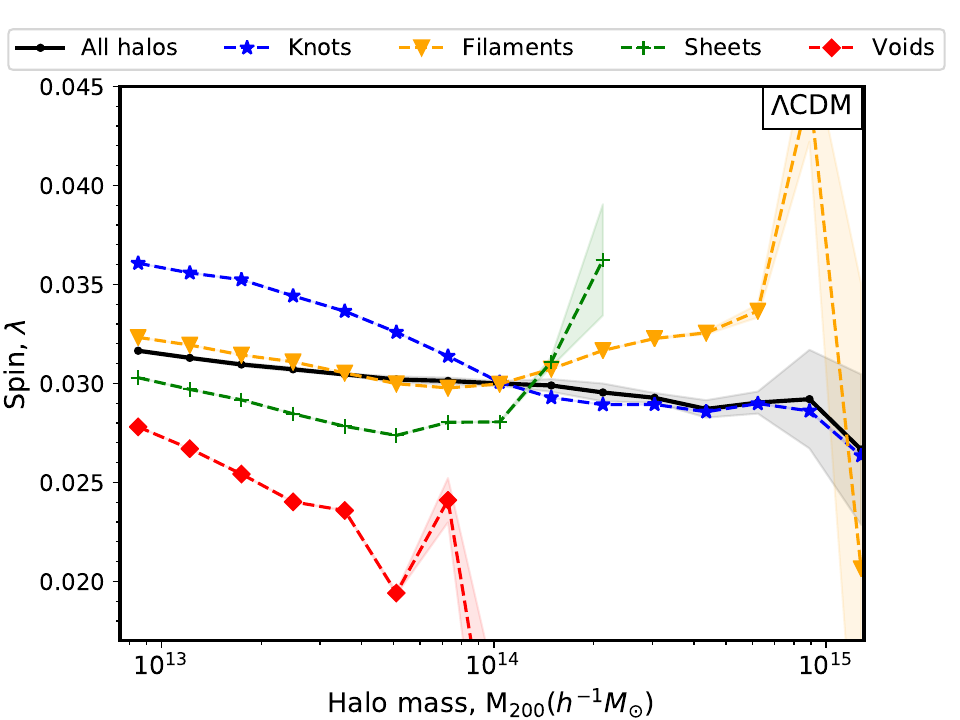}
    \includegraphics[width=0.495\textwidth]{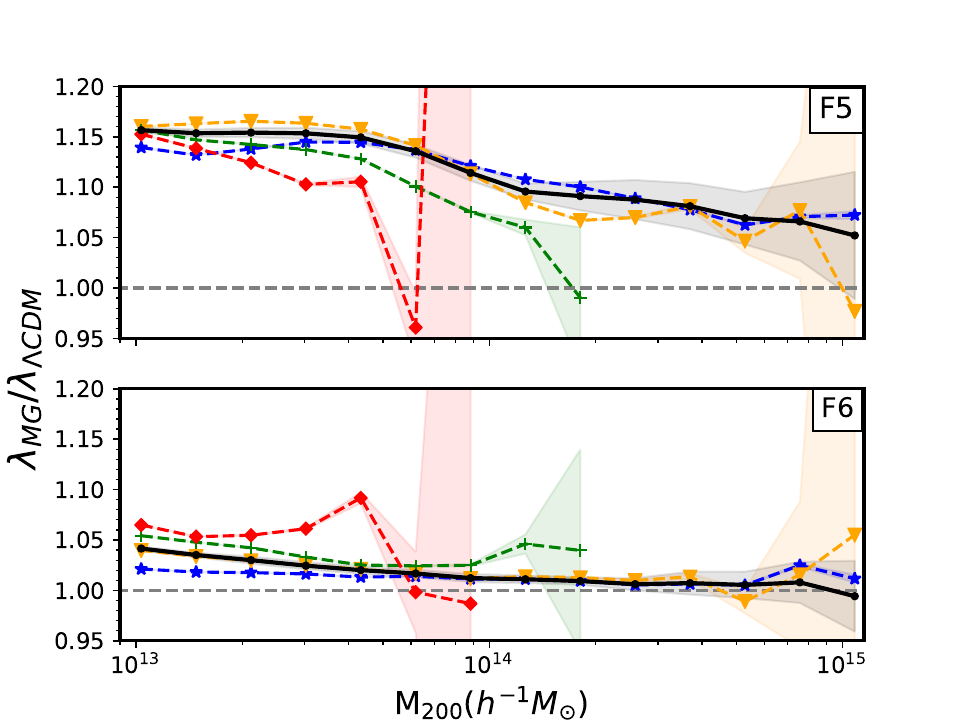}
    \includegraphics[width=0.495\textwidth]{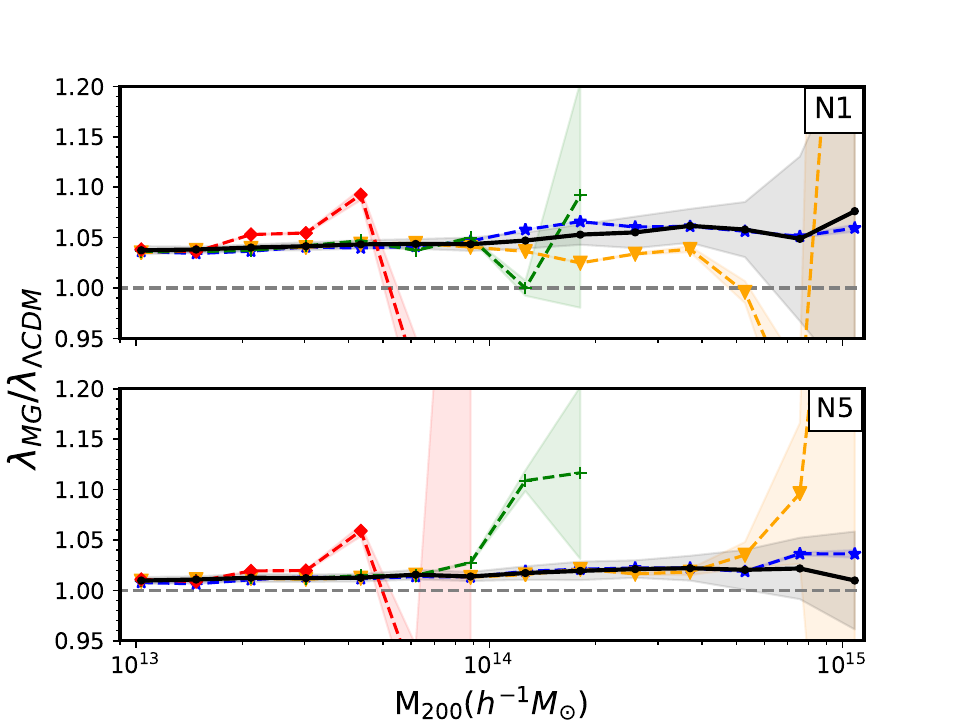}

\caption{\textit{Top plot:} Halo spin, $\lambda$ (\cref{eqn:spin_bullock}) for \lcdm{}, expressed as a function of halo mass, $M_{200}$ (\msunh{}), across each CW environment. Here, the solid black lines represent the $\lambda$ obtained for the entire halo populations, and the dashed lines of different colours are the results for halos separately in each CW environment. The errors correspond to the standard deviation across five realisations. \textit{Bottom plots:} Ratio of the the halo spin between different MG models to \lcdm{}, as a function of halo mass, $M_{200}$ (\msunh{}), for the entire halo population (black lines), and for halos residing in different CW environments (coloured dashed lines). The left plots are the ratio for the $f(R)$ variants, and the right plots are the nDGP ratio results. Here, the shaded regions illustrate propagated errors obtained from the standard deviation across five realisations.}

    \label{fig:spin_all}
\end{figure*}

The CW is a quasi-linear manifestation of the same tidal fields that torque up the protohalo, and are responsible for inducing angular momentum in halos. As a result, we expect that the spin of halos would also correlate with the hosting CW environment. This has already been established in some previous works \citep{hahn_cw_2007,conc_spin_shape_CW,hellwing_CW,cosmic_ballet_3}, where the authors have shown significant influence of the CW environment on the halo spin. In this sub-section, we explore how differently the CW impacts the influence of modifying the underlying gravity theory on the halo spin.

We study the collective impact of MG and CW on the halo spin in \cref{fig:spin_all}. The topmost plot in this figure shows the halo spin across the entire halo mass range probed in this work. Here, we show this result only for \lcdm{}, as MG models exhibit analogous trends. The solid black line in this plot corresponds to the spin for the entire halo population. Furthermore, the dashed lines with different colours are the spins of halos in different CW environments. Clearly, the rotational properties of halos are impacted by the host environment. Namely, knots host the highest spinning halos for $M_{200} \leq 10^{14}$ \msunh{}, and at larger halo masses, the highest spinning halos reside in filaments. At these large halo mass scales, knots trace the spin of the overall halo population. Sheets and voids host the lowest spinning halos at the intermediate halo mass scales (between $7.8 \times 10^{12}-10^{14}$ \msunh{}). This trend is expected as the large halos are predominantly present in the high dense knots region (top plot of \cref{fig:hmf_all}), and therefore a selection of high mass halos is going to follow the distribution of halos in knots. The same is true for the filaments at lower masses.

To study the impact of MG on halo spin across halo mass range, we assess the varying effects of MG on halo spin across each CW environment, as a function of the halo mass in the bottom plots of \cref{fig:spin_all}. Here, the left plots correspond to $f(R)$ results, while the right plots are for the nDGP variants.

The black lines are the ratio of spin between MG and \lcdm{} for all halos which have been discussed in \cref{subsubsec:MG_spin}. Here, we specifically focus on the coloured dashed lines which highlight the ratio separately in each CW environment. For the $f(R)$ gravity, we can see a small environmental dependent trend in the ratio of $f(R)$ to \lcdm{} spin. For masses $\approx 10^{13}$\msunh{}, filament halos have the maximum spin enhancement, and knots have the least enhancement. For F6, voids have the maximum enhancement at halo mass $\approx 10^{13}$\msunh{} ($\approx$ 7\%), and knots have the least enhancement in these mass regimes ($\approx$ 2\%). As halo mass increases, knots follow the ratio trend of all halos at larger masses. This trend in the weaker F6 variant can be attributed to the dominance of chameleon screening in the high-density regions. Here, halos in filaments follow the trend of overall halo samples across the entire halo mass range. For both the $f(R)$ variants, we can clearly see that at large halo masses, the self-screening is more significant which decreases the enhancement in $f(R)$ spin $w.r.t.$ \lcdm{}. This suggests an interesting interplay between the effect of the environment, halo mass and the strength of the \textit{fifth-force} on the halo spin. The environment-dependent chameleon screening impacts the torquing of halos in each environment differently, giving rise to small signatures in the spin. This analysis offers some additional information than what is obtained from the entire halo populations only. 

Contrary to the $f(R)$ results, we do not see an environmental dependent enhancement trend in nDGP to \lcdm{} spin ratio. Halos in all CW environment have similar enhancement as is shown by the overall halo sample. This can be attributed to the scale-independence of the nDGP \textit{fifth-force}. Also, the Vainshtein screening in nDGP depends only on the enclosed halo mass, and not on the surrounding density. This shows that the studies of halo spin in nDGP models for different CW environment do not reveal additional information than the previous studies pertaining to all halo populations. 

% The analysis of MG halo spin in different CW environments highlights that investigating this intrinsic halo property serves as a complimentary test-bed to search for MG signatures. Both $f(R)$ and nDGP exhibit distinct effects on halo spin, with $f(R)$ displaying a notably greater sensitivity to redshift, halo mass, and the CW environmental factors in this analysis compared to nDGP. Therefore, the studies related to this property have the potential to constrain MG. Furthermore, the impact of these modifications to gravity on halo spin are expected to extend to the galaxies hosted within these halos. Previous studies have demonstrated the influence of the CW environment on galactic spins \citep{cosmic_ballet_2,cw_gal,mw_cw,mg_spin_fR}. Thus, these analysis offer a potential to search for the \textit{fifth-force} in the observational domains.

The analysis of MG halo spin in different CW environments shows that halo spin serves as a complimentary, albeit weak, test-bed to search for MG signatures. Both $f(R)$ and nDGP exhibit distinct effects on halo spin, with $f(R)$ displaying a notably greater sensitivity to redshift, halo mass, and the CW environmental factors in this analysis compared to nDGP. The impact of these modifications to gravity on halo spin are expected to extend to the galaxies hosted within these halos. Previous studies have demonstrated the influence of the CW environment on galactic spins \citep{cosmic_ballet_2,cw_gal,mw_cw,mg_spin_fR}. Thus, these results offer a potential to search for the \textit{fifth-force} in the observational domains.

\section{Summary and Discussion}
\label{sec:summary_discussions}

Since the existence of the cosmic web (CW) is a consequence of the intrinsic anisotropic nature of gravitational collapse of the density field, we expect that any modifications to the underlying gravity theory will influence the impact of CW on Large-Scale Structure (LSS) properties. In this context, we investigate how modified gravity (MG) phenomenologies affect dark matter (DM) density fields and halo properties across CW environments. These beyond-GR phenomemologies incorporate additional degrees of freedom in the Einstein-Hilbert action, which modify the perturbation equations governing the structure formation dynamics, and further result in interesting effects on the measures and observables associated with LSS studies \citep{Schmidt_2009_fR, Schmidt_2009_ndgp, MG_SIGNATURES_HIERARCHICAL_CLUSTERING, hmf_sg, drozda_2022, pkmg_sg}. %\spcc{Should you cite this heavily in the conclusion? Maybe best to keep this in the introduction. The text is great, but maybe the citations are unnecessary.} 
In order to quantitatively understand the differences in these large-scale measures from both standard and non-standard gravity theories, we utilize data from the \elephant{} suite of DM-only N-body simulations which incorporates GR and selected MG models: namely Hu-Sawicki $f(R)$ gravity \cite{HS_fR_2007}, and the normal branch of Davli-Gabadadze-Porrati (nDGP) models \cite{ndgp_2000}. Additionally, we use the \tweb{} approach of computing the Hessian of the gravitational potential, in order to divide the density fields from the simulation into four different CW environments: knots, filaments, sheets, and voids. Knots contain the regions of highest densities, which are followed by filaments, sheets and then voids, which are the most underdense regions.

We begin by studying the density distribution in each CW environment, and for each cosmology. All the cosmologies follow a similar log-normal distribution \cite{LN_coles_jones}, with quantitative differences in the median and in the peak of the density distribution across the models. When compared to \lcdm{}, the peak in MG models is shifted towards lower densities. The shift is greater in the stronger F5 and N1 variants, compared to their respective weaker counterparts F6 and N5.

Furthermore, the mean of the overall density in each model is normalised to unity. However, on comparing the mean of densities across CW environments, knots tend to become more dense as the strength of gravity increases, and the mean in filaments for all models coincides with \lcdm{}. On the contrary, sheets and voids become less dense. This suggests an enhanced effective transfer of matter from voids to knots via sheets and filaments.

The late-time non-linear gravitational evolution causes departure from Gaussianity, which makes the study of higher-order clustering relevant. Therefore it is valuable to have robust measurement of these statistics and their associated uncertainties in order to obtain accurate cosmological parameter estimations and constraints on the MG models. Here, we probe the higher-order clustering moments to quantitatively differentiate the density distribution across cosmologies. In this work, we investigate the two-point variance and higher-order statistics of reduced skewness and kurtosis.  We first examine the impact of the CW environment on these statistics within the \lcdm{} framework and subsequently explore the role of MG in influencing these statistics across different environments. We summarise the findings of this sub-section in the following points:
\begin{itemize}

\item Variance, reduced skewness and reduced kurtosis show a clear dependence on the CW environment, which is most pronounced on small scales. The results for all environments converge to the trend for the overall density on the larger scales. This convergence on large scales is expected since we smooth more fluctuations as we increase the smoothing radii. Also, smoothing the density on larger and larger scales converges the density of all the environments towards the overall mean density.

\item On comparing the MG to \lcdm{} variance, we find that both $f(R)$ and nDGP have significantly different impact on the variance. $f(R)$ variants show a scale-dependent trend in the ratio on smaller scales, with the results of all the environments approaching unity on the larger scales. While, for nDGP, we notice a similar scale and environmental dependent ratio on small scales, with all results converging to a constant enhancement on the large scales. 

\item We further compare the MG and \lcdm{} reduced skewness. The results between $f(R)$ and nDGP are again quantitatively different. All environments on the small-scales exhibit different trends, which converge to unity as we approach larger scales. Contrary to variance, the reduced skewness in the overall density for $f(R)$ variants decreases when compared to \lcdm{}. In all cases, voids in MG depart the most from \lcdm{}.

\item The findings for reduced kurtosis exhibit similarities to reduced skewness. However, in the context of reduced kurtosis, MG models demonstrate a more pronounced departure from \lcdm{} compared to the reduced skewness results.
%\spcc{Worth noting that, in all cases, voids are the CW environment that depart from LCDM the most in the MG cosmologies.}

\end{itemize}

From these results, we confirm the potential of studying the hierarchical amplitudes as potential probes to quantify the modified gravitational dynamics, a finding already emphasised in clustering studies for a wider class of MG theories \cite{hierarchical_rebel,hierarchical_fR,MG_SIGNATURES_HIERARCHICAL_CLUSTERING,drozda_2022}. Furthermore, we showcase the importance of probing these statistics individually in each CW environment. The additional information encoded in the environmental dependence of these higher-order statistics can further be used to break degeneracies present in the measurements of the clustering statistics when averaged across all environments. This would in turn help forecast better constraints on the cosmological quantities that we obtain from the studies of higher-order clustering \cite{PNG_MOMENTS,png_moments_2,cw_pk_1,cw_pk_2,juszkiewicz_skewness_bao}, and further aim for better constraints on the theory of gravity.

Moving to the halo properties, we first focus on the Halo Mass Function (HMF) within our MG models for different CW environments. Notably, the HMF demonstrates sensitivity to the specific environment hosting the halo, an already established result \citep{hahn_cw_2007,halo_abundance_cw,vweb_yehuda,evolution_cw_nexus,hmf_fR_cw,tracing_cw_libeskind_2018,hellwing_CW}. Additionally, upon a comparative analysis of the MG impact on HMF, we observe the influencing role of the CW in shaping the effects of MG on HMF. %\spcc{Not sure about this last sentence.}

Here, we summarise our findings for the HMF section:
\begin{itemize}
    \item The HMF trend shows an explicit dependence on the CW environment, with the fractions of halos residing in different CW environments varying as a function of their mass. For the mass range probed in this work, we see that the filaments host the majority of halos for $M < 10^{14}$ \msunh{}, and higher mass halos mostly reside in knots. This trend is similar for both \lcdm{} and MG models.
    \item We further study the impact of MG on HMF in different CW environments. Both $f(R)$ and nDGP show quantitatively different trends across halo masses and environments. The ratio for the overall HMF shows a peak-like feature in $f(R)$ and a monotonic enhancement in nDGP. This was already discussed in \cite{hmf_sg}. The environmental-dependent trend across MG variants is more evident for halos with $M < 10^{14}$ \msunh{}, and the ratio converges to the overall halo population at large halo masses. This results from an interplay between the influence of the \textit{fifth-force}, and the physics of the screening mechanism manifesting across halo mass ranges. 
\end{itemize}

The second halo property we probe in this work is the halo spin. The analysis of the spin distribution for the entire halo population reveals that, like \lcdm{}, both $f(R)$ and nDGP are characterised by log-normal distribution. However, they exhibit a higher mean spin parameter, $\lambda_0$, and a greater standard deviation, $\sigma_\lambda$. This shows that the \textit{fifth-force} speeds up the halos and results in greater spin in these MG scenarios compared to \lcdm{}. The enhancement in both models though is phenomenologically different. Upon deeper investigation, we find that the evolution of spin enhancement in MG models across redshifts has a direct dependence on the change in growth rate, $\dot{D(t)}$. $f(R)$ variants show both mass and redshift-dependent enhancement in the spin, whereas nDGP variants show redshift independent, though a weakly mass-dependent spin enhancement. Moreover, the spin enhancement in the stronger F5 model reaches up to $\approx$ 15\% for $M \approx 10^{13}$ \msunh{}, whereas enhancement in N1 lies between $5-10$\% across the entire halo mass range studied here.

We further study the impact of CW on halo spin and our findings are summarised in the following points:
\begin{itemize}
    \item Halo spin shows an explicit dependence on the environment hosting the halo, with the highest spinning halos mostly residing in the dense environments of filaments and knots. We can see a clear dependence on the environment for halo masses $\leq 10^{14}$ \msunh{}. Other cosmologies show similar results.
    \item On probing the impact of MG on halo spin in different CW environments, we find that both $f(R)$ and nDGP have a different impact on the spin. $f(R)$ models show enhanced spin, with a more pronounced halo mass and weak environmental dependence. On the other hand, the \textit{fifth-force} in nDGP does not have significant impact on the spin, and this difference from \lcdm{} is also independent of the CW environment. 
\end{itemize}

The results from analysing the halo properties highlight that CW has a strong influence on the environment hosting the halos. This result has already been shown for \lcdm{} but is extend to MG phenomenologies in this work. The abundance and rotation of halos increase in MG when compared to \lcdm{} where enhancement can be a strong function of the hosting environment. Since halos form the site of galaxy formation, these differences are expected to also propagate in the studies of galactic and cluster dynamics.

Our study focuses on DM density field and DM halo properties. The large-scale DM distribution, together with the halos aggregated within form the first layer of a theoretical framework upon which we build further our understanding of galaxy formation and evolution. All the effects we found and listed above are thus connected to this base-ground level theoretical framework. The natural consequence of this is that none of the MG effects we elucidate in the CW of DM can be directly translated to astronomical observables. However, our work establishes and highlights the significant differences that can, and should be taken into account when one is modeling various observables that concern the LSS and the CW. For instance, the MG signal encoded in the higher moments of the DM density field can be translated into more specific predictions for weak lensing statistics such as shear and convergence power spectra \cite{darkside_WL,wl_mg_Schmidt_2008,wl_predictions_nl_pk,cosmological_tests_gravity_martinelli_casas}. If the lensing statistics can be robustly further split across the CW environments, then the expected net MG effects will be different in different web segments. The environmental effects induced on the HMF indicate, that if a halo mass to stellar mass relation can be modeled for MG, with a precision already achieved for \lcdm{}, then this promises a potential for a strong prediction of different abundance of high stellar mass galaxies in voids between MG and \lcdm{} \citep{fR_void_1,ndgp_void_2}. Finally, the halo spin signal we see across environments and MG models has potential to materialize as an effect in angular momentum, and inturn on the spiral galaxy formation efficiency in \lcdm{} and MG \citep{bird_spin_2020,welker_spin_2020,obreja_spin_2022,ansar_spin_2023}. Translating and connecting our DM-based theoretical level prediction to the language of galaxy observations is out of the scope of our work here. It will be an exciting and natural next step, which we leave for future work.

This work highlights that incorporating the morphological classification of the CW into the existing analysis of DM density and halo properties increases their utility as a probe of MG. Every property investigated in this study demonstrates a nuanced dependence on its hosting environment. Additionally, our study reveals distinct MG effects across various environments. This emphasizes the value of CW studies in MG, providing insights into the dynamics of the $\textit{fifth-force}$ and screening mechanisms. Our environment-dependent investigation yields richer information compared to the overall analysis averaged across all environments. Consequently, constraints derived from combining LSS properties from each specific environment will potentially be more constraining than those obtained from an analysis of the entire density field.

\begin{acknowledgments}
The authors gratefully acknowledge the insightful discussions with Maciej Bilicki, which significantly influenced the results of the manuscript. SG and WAH are supported via the research project ``VErTIGO'' funded by the National Science Center, Poland, under agreement no. 2018/30/E/ST9/00698. This project also benefited from numerical computations performed at the Interdisciplinary Centre for Mathematical and Computational Modeling (ICM), the University of Warsaw under grants no GA67-17 and  GB79-7. SG also acknowledges partial support from the Young Researcher's Grant 2022, CTP PAS. SP acknowledges financial support from the Deutsche Forschungs Gemeinschaft joint Polish-German research project LI 2015/7-1. PGV is supported by the research grant ‘Optimizing the extraction of cosmological information from Large Scale Structure analysis in view of the next large spectroscopic surveys’ from MIUR, PRIN 2022 (grant 2022NY2ZRS 001).
\end{acknowledgments}
\bibliography{MG_CW}% Produces the bibliography via BibTeX.

\end{document}